\documentclass[aps,twocolumn]{revtex4-1}
\usepackage{amsmath,amssymb}
\usepackage{slashed}
\usepackage{graphicx}
\usepackage{bm}
\usepackage[T1]{fontenc}
\usepackage[utf8]{inputenc}
\usepackage{gauss} 
\usepackage[top=1.0in,bottom=1.0in,left=1.0in,right=1.0in]{geometry}
\usepackage[colorlinks,linkcolor=blue,citecolor=blue]{hyperref}
\newcommand{\be}{\begin{equation}}
\newcommand{\ee}{\end{equation}}
\newcommand{\bea}{\begin{eqnarray}}
\newcommand{\eea}{\end{eqnarray}}
\newcommand{\ba}{\begin{eqnarray}}
\newcommand{\ea}{\end{eqnarray}}

\begin{document}

\title{Baryons and tetraquarks using instanton-induced interactions
}

\author{Nicholas Miesch}
\email{nicholas.miesch@stonybrook.edu}
\affiliation{Center for Nuclear Theory, Department of Physics and Astronomy, Stony Brook University, Stony Brook, New York 11794--3800, USA}

\author{Edward Shuryak}
\email{edward.shuryak@stonybrook.edu}
\affiliation{Center for Nuclear Theory, Department of Physics and Astronomy, Stony Brook University, Stony Brook, New York 11794--3800, USA}

\author{Ismail Zahed}
\email{ismail.zahed@stonybrook.edu}
\affiliation{Center for Nuclear Theory, Department of Physics and Astronomy, Stony Brook University, Stony Brook, New York 11794--3800, USA}

\begin{abstract} 
We  analyze some aspects of the
perturbative and non-perturbative interactions  in the composition
of heavy quarkonia, heavy and light baryons ($ccc$ and $uuu$ ones), as well
as all charm tetraquarks ($cc\bar c\bar c$). 
Using the hyper-spherical approximation
and effective radial potentials (in 6 and 9 dimensions, respectively) we derive
their spectra and wave functions. In all of the cases, we  focus on the splittings between the
s-shell levels, which are remarkably insensitive to the quark masses, but proportional to
the effective interaction potentials. We use the traditional Cornell-like potentials, and the
non-perturbative instanton-induced static potentials,  from correlators of two, three and four Wilson lines, and find rather satisfactory description of spectra in all cases. 
 \end{abstract}

\maketitle
\section{Introduction}
One of the goals  of the present paper 
is to reassess the information one can extract from
hadronic spectroscopy, on quark-quark and quark-antiquark interactions
in multi-quark system, baryons and tetraquarks. Of course, we do not
attempt to cover this vast field, but rather focus on some
particular (flavor-symmetric) objects. Furthermore, we focus on
specific features of spectroscopy that are most sensitive to the inter-quark interactions (and  less sensitive to the quark masses and additive constants in the potentials).

As examples of these observables,  we use the {\em splittings} between the  $1S,2S,3S$ states ( S-shell, as usual means zero angular momentum), and evaluate the splittings between the p-shell states (negative parity).  
We start  in section \ref{sec_pot_QbarQ} with a very traditional issue,
in relation to the central quarkonium potential in charmonium and bottomonium. 
In spite of the large mass difference between $c,b$ quarks, these splittings in are very similar. While this fact by itself is widely known, we elaborate in more detail on its sensitivity to the exact shape of the potentials. More specifically, we use comparatively 
the Cornell (Coulomb-plus-linear) potential, the Martin potential, and
the instanton-induced potential we derived in~\cite{Shuryak:2021fsu} (our 
first paper of the light-front series) .

In sections \ref{sec_inst_meson} and \ref{sec_hyper_baryons} we
discuss the flavor-symmetric baryons
($ccc,uuu$)  and $cc\bar c \bar c$ tetraquarks, respectively. As a method,
we use the so called {\em hyperdistance} (or hypercentral, or K-harmonics) approximation, by reducing quantum mechanics (in 6 dimensions for baryons and 9 for tetraquarks) to a single radial Schrodinger equation. The method
has been developed in the 1960's, in nuclear physics, for the quantum mechanical treatment of light nuclei, see e.g. \cite{Simonov:1965ei}.
One well-known four-body example is  $^4He$, for which
   this approach  reproduces  its binding $\approx 28\, MeV$ from
  conventional nuclear forces. Furthermore, in a
relatively recent paper \cite{Shuryak:2018lgd} on 4-nucleon clustering 
in heavy ion collisions at finite temperatures,  
 it was found that the same hyperdistance equation also correctly predicts the $second$ (2S) level of $^4He$, with small binding and close to its experimentally known location. This work increased
our trust in this approach, at least for the splittings of the s-shell states.

 The applications 
of the hyperdistance approximation to various multiquark states 
have also been done over the years,  see e.g.
\cite{Badalian:1985es,Chernyshev:1995gj,Ferraris:1995ui,Gandhi:2018lez,Brambilla:2022ura} and references therein.
With the recent revival of hadronic spectroscopy due to the discovery of multiple
tetra- and penta-quark states, there has been  renewed interest
in few-body quantum mechanics in general, and in the hyperdistance approximation in particular.

In sections \ref{sec_hyper_baryons} and \ref{sec_hyper_tetra}, we start
our analyses using Cornell-like potentials,  and focus chiefly on 
the level splittings of the  S-shell states. Much like in  quarkonia,
we found that the splittings are insensitive to the quark mass values:
for $ccc$ and $uuu$ baryons, 
these splittings are nearly identical, see Fig.\ref{fig_ccc}. What they are
sensitive to is the overall strength of the confining force 
in multiquark systems, to which they are roughly proportional. 
(This is highly
nontrivial, since the wave functions and even sizes of these systems are vastly different!)
  
The main physics issues this paper is aimed at are the following questions:
Can the interactions inside few-body hadrons (such as baryons and tetraquarks specifically) be approximated by a sum of two-body ones? 
Do they only
depend  on the {\it hyper-distance} variable (related to the sum of  distances between all constituents squared)? If so, are the shapes of the corresponding effective
potentials similar to the Coulomb+linear potentials used for quarkonia?
What are the magnitude of their respective coefficients, for 3 and 4 quarks? 
What is the order of magnitude of the  corrections stemming from  the lowest hyper-spherical approximation? 

Static interquark potentials have well known expressions in terms of
the correlators of Wilson lines. 
Lattice gauge theories in fact started with the evaluation  of 
correlated Wilson lines $\langle W W^\dagger\rangle $,  
and the shape
of the $\bar q q$ static potential.  The literature on this subject is vast and cannot
be covered in this work. Yet the lattice studies of correlators of three or four
Wilson lines are surprisingly limited, with results limited to simple geometries for
forces in baryons  \cite{Koma:2017hcm} 
and in tetraquarks \cite{1702.07789}. For the baryons, three quarks are set
at the corners of triangles of different shapes and sizes, while for tetraquarks four quarks are set at the corners of
rectangles $r\times R$ with variable $r,R$. Unfortunately, no attempts to project
 the results on the hyperdistance variable were made so far.

While numerical lattice simulations are first-principle-based approaches to vacuum ensemble of gauge fields,  some simple models for the vacuum have been developed over the years. Instantons are 4-d spherical solutions in Euclidean gauge theory vacua,
describing tunneling between configurations with different Chern-Simons numbers. They are the basis of semiclassical approximation and trans-series
in QFT. Here, we will not go into the theoretical details of the model, 
but just just state that we will
use what we refer to as the  ``dense instanton vacuum model". 
Unlike earlier version, the
``dilute  instanton vacuum" \cite{Shuryak:1981ff} (which included only isolated instantons and
focused on their fermionic zero modes and chiral symmetry breaking, for review see \cite{Schafer:1996wv}), it also includes close instanton-antiinstanton pairs which contribute to Wilson lines by
their fields. The two parameters of the model are the 4d density $n_{I+\bar I}\approx 7 \, {\rm fm}^{-4}$
of instantons and antiinstantons, and their mean size $\rho\approx \frac 13 \, {\rm fm}$.
For discussion of these parameters and its relation to static quarkonium potential see
\cite{Shuryak:2021fsu}. Throughout the same parameters will be used in the evaluation of the effective potentials in all systems considered, e.g. $\bar q q, qqq, qq\bar q\bar q$. So the resulting potentials  are
essentially parameter free. All we do is to generalize the calculation of Wilson line
correlators $\langle W W^\dagger \rangle $, to 
$WWW$ and $WWW^\dagger W^\dagger$ pertinent correlators.
We are satisfied to find that the resulting potentials do indeed reproduce
the pertinent hadron spectra in all cases.

\section{Central potential in quarkonia} 
\label{sec_pot_QbarQ}
Since the early 1970's, when heavy $c,b$ quarks were discovered, the nonrelativistic description
of quarkonia  $\bar c c, \bar b b$ states became the main pillars of hadronic spectroscopy.
 And yet, we decided to start again from quarkonia, to test
how well the proposed potentials work. 

\begin{figure}[h]
\begin{center}
\includegraphics[width=8cm]{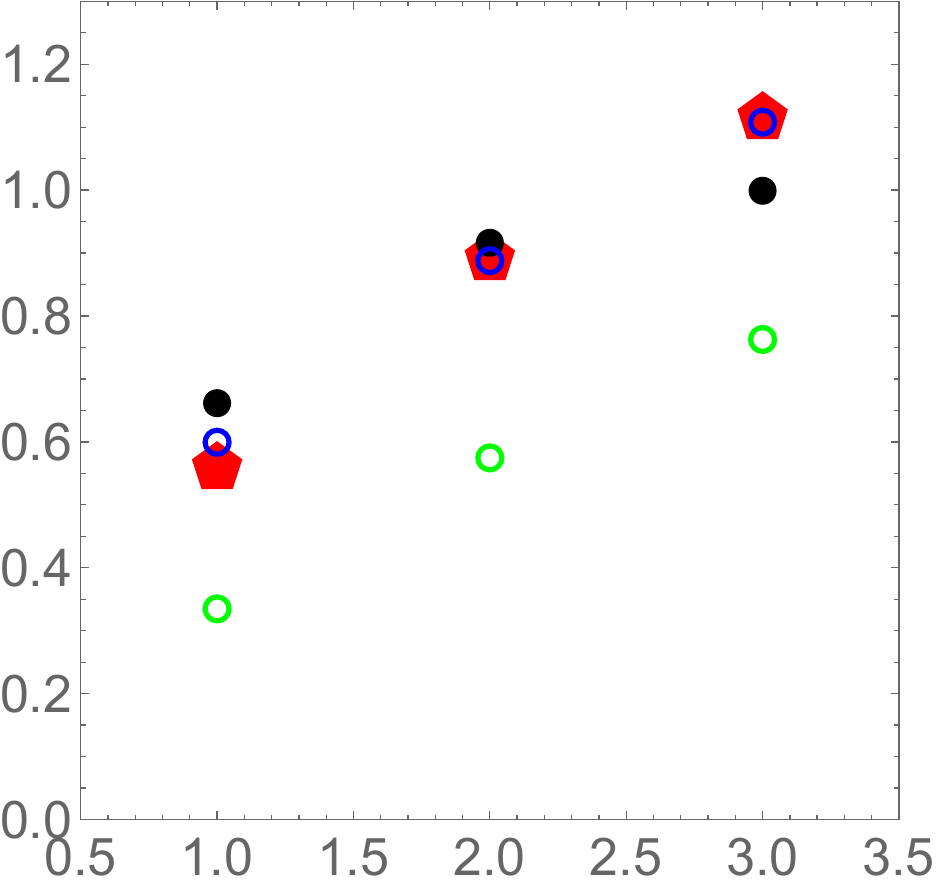}
\includegraphics[width=8cm]{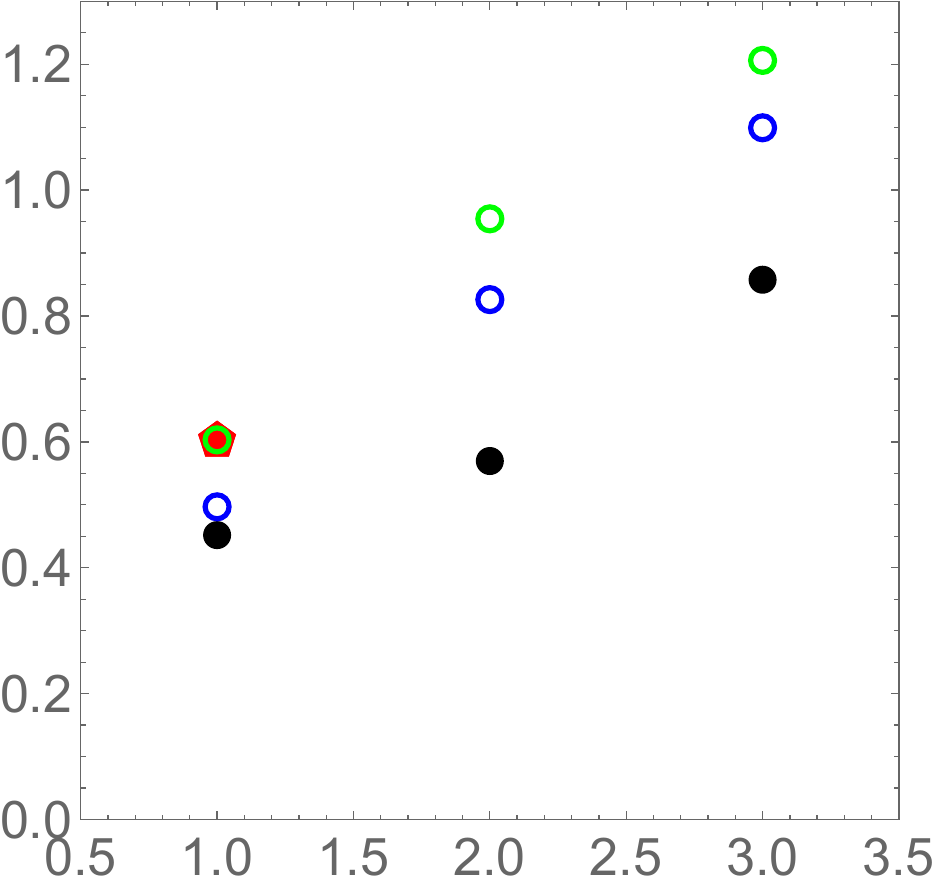}
\caption{Mass splittings in GeV, of the $\Upsilon(nS)-\Upsilon(1S), n=2,3,4$  (upper plots)
and spin-averaged splittings in $\bar c c$ systems (lower plots) shown versus $n-1$ by red pentagons. The blue circles correspond to solution of the Schrodinger equation using the Cornell potentials, the green circles are for the Martin potential, and the black disks are for the instanton-induced potential explained in the text. }
\label{fig_splittings}
\end{center}
\end{figure}

A good motivation is to also  include current data on bottomonium states, as listed in 
the review of particle physics 2022~\cite{ParticleDataGroup:2022pth}. We focus below on the
S-level  $splittings$, insensitive to quark masses and 
 overall constant terms in the potentials. One such input is the splitting between the spin-averaged charmonium masses 
 \be 
 M_{spin-average}=(3M_{\psi}+M_{\eta_c})/4    
 \ee            
in the 2S and 1S shell
\be 
M_{spin-average}^{2S}-M_{spin-average}^{1S}\approx 605\, MeV
\ee
In the bottomonium family we now have $\Upsilon$ states listed up to 4S, but not enough
$\eta_c$ states to perform spin average for all of them. Hence, we will use the Upsilon mass splittings instead. (Spin-dependent forces are not yet included, but
those are $\sim 1/m_b^2$ and rather small in bottomonia.)
These data points are shown as red pentagons in~Fig.\ref{fig_splittings}.

The precise shape of static potential, describing  known quarkonium masses, is an old problem, discussed many times since e.g.  the first reviews such as
\cite{Grosse:1979xm}. 
The classic Cornell potential is the sum of  the one-gluon exchange Coulomb term, and the linear confining term. For definiteness, we use its version with coefficients 
\be \label{eqn_Cornell}
V_{Cornell}= - {0.641\over r} +0.113\cdot  r
\ee
(Throughout, all dimensions are in GeV units, e.g. the distance $r$ is in $GeV^{-1}$, and the string tension in $GeV^{2}$.)
The  two parameters in the potential are fitted so as to reproduce the three known bottomonium splittings $2S-1S,3S-1S,4S-1S$,   shown in the upper plot by the blue circles.
 Note that it does not reproduce  well the charmonium splittings as shown in the lower pannel of Fig.~\ref{fig_splittings}.

Let us remind the reader that potentials which grow as smaller power of $r$, display a smaller growth of energy with $n$. (
Only an oscillator $V\sim r^2$ has $E_n\sim n$ linearly.) 
 Early studies argued 
that static potentials $V\sim r^\alpha $ with $\alpha<1$ 
can describe spectra better. The so called {\em Martin potential}  fitted to (then available) data, see e.g.   \cite{Grosse:1979xm} 
\be V_{Martin}(r)= - 8.054+6.870\cdot r^{0.1} \ee
(all in GeV) has a very small power, and therefore is close to a logarithmic potential, for which it is known that
the 1S-2S splittings of the charmonium and bottomonium systems would be equal.
And indeed,  they are almost the same. More specifically, the spin-weighted average (1S, 2S) masses are (3068.65, 3673.95) MeV for charmonium and (9444.9, 10017.2) MeV for bottomonium, so M(2S)-M(1S) = (605.3,572.3) MeV for ($c\bar c, b\bar b$). 

As one can see from the lower pannel of Fig.\ref{fig_splittings}, with the  Martin potential predictions for splittings are shown by green circles, it indeed does give a
very accurate number for the first charmonium splitting (to which  the 
potential parameter was originally  fitted). 
Yet its predictions  
for $\bar b b$ Upsilon states (upper plot) is rather far from the data points.

To summarize  our preliminary discussion:
the Cornell potential is still perhaps the best choice, reproducing well  these particular observables both for  $\bar c c$ and $\bar b b$ systems. The
potentials with smaller power, such as the
Martin potential, do fit charmonium data but are not very good for bottomonia.

\subsection{Instanton-induced forces in mesons} 
\label{sec_inst_meson}

Understanding  the hadronic spectroscopy can be viewed as only a step 
towards the even more ambitious goal of understanding the structure
of the (Euclidean) gauge fields in the {\em QCD vacuum}.  It may be approximate and schematic, yet it allows for practical evaluation of all kinds of
{\em non-local correlations functions} of nonperturbative  gauge fields, 
 without supercomputers. The static potentials are the nonlocal
 correlators of fields inside the Wilson lines, the
 path-ordered exponents of color fields.
 
 Instantons are nontrivial gauge configurations, being the basis
 of the semiclassical theory. Their properties in the QCD vacuum, chiral symmetry breaking and 
 physical effects on mesonic and baryonic point-to-point correlators
 were reviewed in \cite{Schafer:1996wv}. Recent discussion including the role of instanton-antiinstanton configurations on the lattice and for effective potentials can be found in
 the first paper of this series \cite{Shuryak:2021fsu}.

The contribution of instantons to the quark potential is an old subject, pioneered by  \cite{Callan:1978ye}, but its practical
application to central and spin-dependent forces between quarks 
attracted very little attention~\cite{Chernyshev:1995gj}. In
 our recent series of papers \cite{Shuryak:2021fsu,Shuryak:2021hng,Shuryak:2021mlh}, a novel version of the instanton model was proposed for that. It incorporates lattice data on close instanton-antiinstanton pairs and was shown
 to generate  the  $\bar q q$ static potential which, while not confining per se,
 is close to the Cornell potential
 up to about 1 fm distances. 
 
 We then extended this model to calculation of the spin-dependent $\bar q q$ forces,
 and static quark-quark potentials in diquarks and baryons. Those
 can be expressed as Wilson lines, decorated  by two field strengths. In the cases of spin-spin and tensor forces, those fields are gluo-magnetic. 
 (Note that attributing confinement to
  gluo-electric flux tubes, one is faced with the  problematic of the spin-dependent
  forces. Indeed, there is no place as such for magnetic fields, and the only
 contributution is  via the Thomas' precession. The instantons are self-dual,
 with gluomagnetic and gluoelectric components being equal, and thus they
do generate significant spin-dependent forces.)

We now briefly recall some general points in the derivation of
the $\bar q q$ potentials. The Wilson lines go along the
world-path of static charges along the Euclidean time. The produced $W$
operators are
unitary color matrices, describing  rotations of quark color vectors by certain 
angles and in certain directions, around which these rotations are to be made. Let
 the 3-vectors representing distances from the quark locations $\vec r_i, i=1...N$ to the instanton center be $\vec \gamma_i=\vec r_i - \vec y$ (amusingly it acts as a {\it vertex} center).
The color rotation angles are 
\be \label{eqn_angles}
\alpha_i=\pi\bigg(1- {\gamma_i \over \sqrt{\gamma_i^2+\rho^2}}\bigg) \ee
where $\rho$ is the instanton size.
Far from the instanton, $\gamma_i \gg \rho $, this angle vanishes. At the instanton center $\gamma_i=0$, the angle is maximal $\alpha_i=\pi$.
 The  $SU(2)$ part of the Wilson line can be expressed using Pauli (rather than Gell-Mann) matrices $\tau^A,A=1,2,3$
\bea
{\bf W}^a_{lb}&=&{\rm exp}(\pm i \vec \tau\cdot \vec n_l \alpha_l)\\
&=& 
\big(cos(\alpha_l)\,{\bf \hat 1}-i(\vec \tau\cdot \vec n_l)\,sin(\alpha_l)\big)^a_b\nonumber
\eea
where the unit vectors are  $n_i=\vec\gamma_i/\gamma_i$, and $ a,b=1,2; l=1...N_q$.
The third color is not affected by the instanton fields, so this $2\times 2$ matrix 
should be complemented trivially by the unit value in the third row.

The integral over (large) Euclidean time $\tau\in [-T/2,T/2]$ may include some number
of instantons $N_{inst}$. Summing over $N_{inst}$ leads to the exponentiation of the correlator. The exponent is extensive in the Euclidean time
$T$ times the potential 
\be
V(r)=n_{I+\bar I}\int d^3 y  {1 \over N_c} {\rm Tr}\big(\hat {\bf 1} - W(0)W^\dagger(\vec r)\big)
\ee 
For the plot of it see Fig.5 of \cite{Shuryak:2021fsu}.
Below  these formulae will be generalized for 3 and 4 quark lines.
 These calculations can later be straightforwardly generalized to other multi-quark hadrons. We plan to calculate the spin-dependent forces in the same settings in forthcoming publications. 

The instanton-induced potential together with the Coulomb term, can be put
into the Schrodinger equation, to derive the quarkonium states.
Let us now return to Fig.\ref{fig_splittings} for the mass differences between
the excited $2S,3S,4S$ states, from the $1S$ ground state of $\bar b b, \bar c c$ quarkonia. The splittings for the
 instanton-based potential are shown 
 in Fig.\ref{fig_splittings}  by
 black discs.  For the
 2S-1S and 3S-1S splittings of bottomonia the results match the data quite well,
  and only for the 4S state they visibly deviate from the data, by about 10\% or so.
   For charmonia this potential leads to smaller level splittings.   
   Overall, its accuracy is in  between the Cornell and Martin potentials. Note however, that unlike those, fitted to reproduce quarkonia levels,
 the instanton parameters (sizes  and density)  were {\it not fitted} to spectroscopy, 
 the instanton density and mean size
  taken  directly from lattice data about vacuum topology.

\begin{figure}[h]
\begin{center}
\includegraphics[width=4cm]{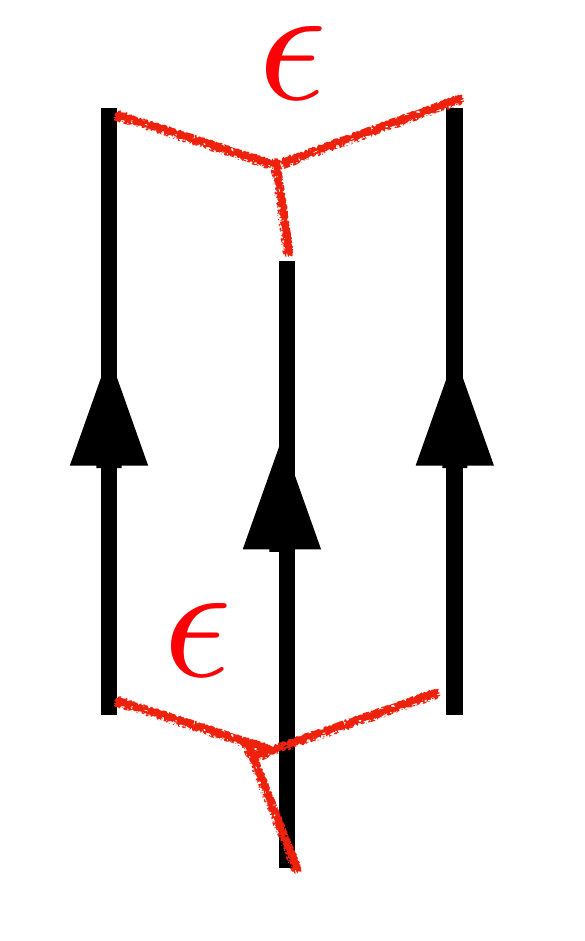}
\caption{Correlator of three Wilson lines in baryons, connected
by a vertex with Levi-Cevita symbol.}
\label{fig_WWW}
\end{center}
\end{figure}

\section{Baryons: hyper-spherical approximation and instanton-induced potential}

\subsection{Baryons in hyperspherical approximation} \label{sec_hyper_baryons}
We started this paper discussing heavy quarkonia, and we will end discussing
full-charm tetraquarks. However, it is also natural to address
heavy-quark baryons, $ccc$ or $bbb$, and anything in between. Unfortunately,
these hadrons  still elude experiments, so we will also include calculations for
baryons composed of light quarks. In doing so, we will assign the same flavor for all three quarks, i.e. $uuu=\Delta^{++}$ states. (There is not enough data on $sss=\Omega^-$ spectra.)

We will use 
the {\em hyper-spherical approximation}, also known as the method of K-harmonics.
 The main idea is that the lowest states
depend mostly on the radial hyperspherical coordinate in 6-d space of Jacobi coordinates
\be R_6^2=\vec \rho^2+\vec \lambda^2
\ee
with a small admixture of components with angular dependences. In standard terms, 
we study 6-d
 s-shell states. Their wave function can be obtained
by solving a single $radial$ Schroedinger equation. The method has been used
for baryons over the years, see e.g. \cite{PhysRevD.24.1343} for heavy baryons of interest.
The kinematical details of this approximation are given in 
Appendix~{\ref{sec_hyper_general},\ref{sec_jacobi_baryons}}.

The radial Schrodinger equation for the $S$ states
has the form 
\be 
\label{RED6}
\bigg(-u"+{15\over 4 R_6^2 }u \bigg){1\over 2M}+(V_6-E_{Ns}) u=0
\ee
where the reduced wave function $u(R_6)$ is related to the radial wave function by 
\be \psi(R_6)=u(R_6)/R_6^{5/2}
\ee
Note, that $R_6^5$ gets absorbed in the volume element, so that $u$ has the usual normalization, as in one dimension $\sim \int dR_6 |u(R_6)|^2$.
This reduction eliminates the first derivative from the Laplacian, but adds  an  extra ``quasi-centrifugal" term. The radial  projection of the potential on $R_6$ is defined in the Appendix~\ref{sec_jacobi_baryons}.

\begin{figure}[h]
\begin{center}
\includegraphics[width=8cm]{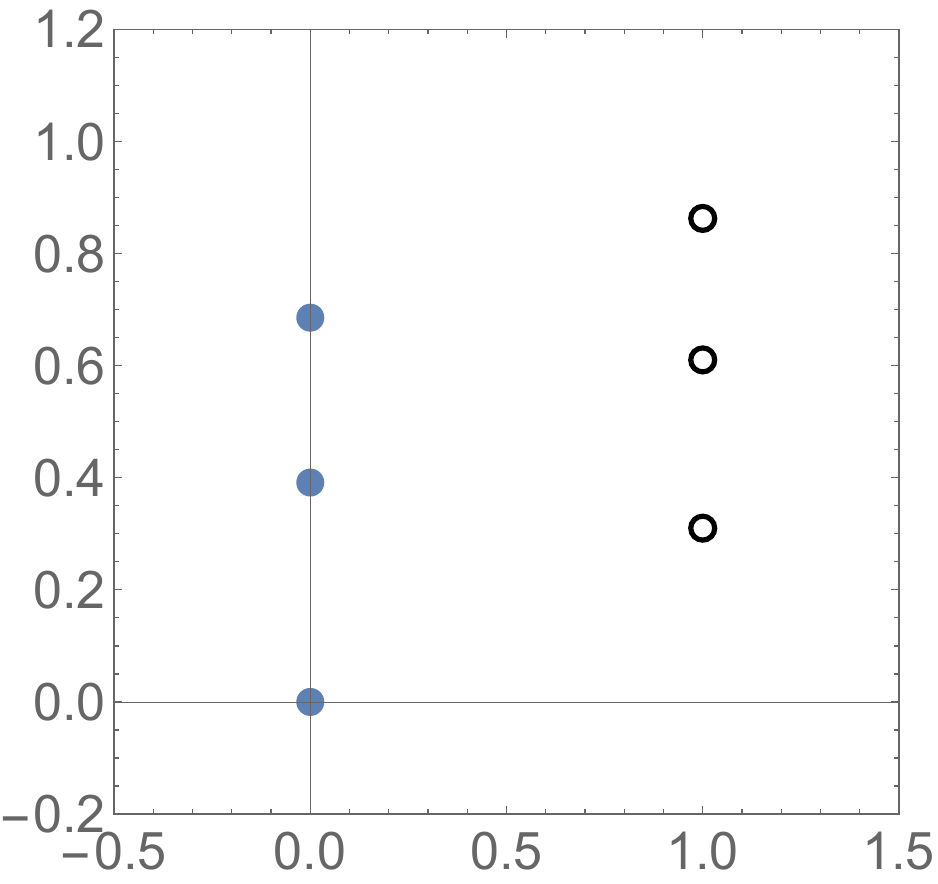}
\includegraphics[width=8cm]{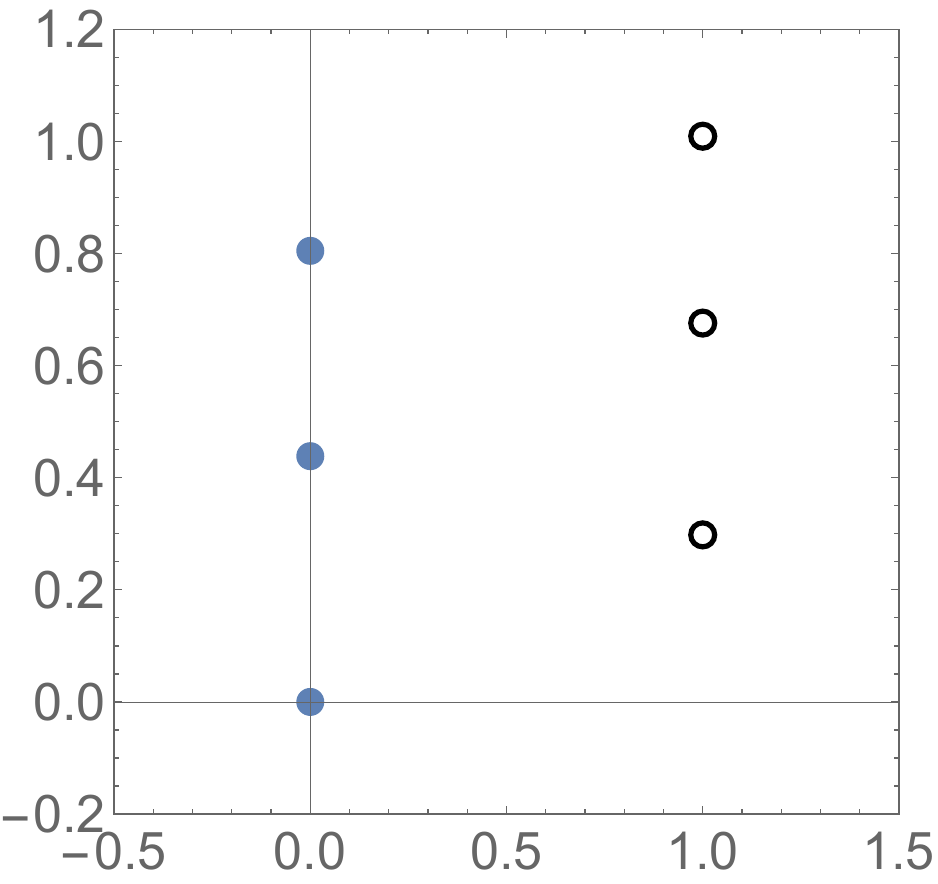}
\caption{Masses of $ccc$ baryons (upper plot) and $uuu=Delta^{++}$ (lower plot)  $M(N,L)$, for $L=0$ and $L=1$ shells, counted from the ground states. }
\label{fig_ccc}
\end{center}
\end{figure}

The mass splittings  of the lowest s- and p-states, $ccc$ and $uuu$ baryons, 
following numerically from (\ref{RED6}) are shown in Fig.\ref{fig_ccc}. The binary $QQ$ potential is taken as  $1/2$ of the Cornell potential (used above for $\Upsilon$
family).  We do not show  the absolute masses but 
focus instead on the level splittings.
 Experience shows that the splittings 
have relatively little dependence on the choice of the quark mass, but
depend primarily on the interaction potential. Indeed, as shown in
Fig.\ref{fig_ccc}, the plots for $ccc$ and $uuu$ baryons look nearly identical, in spite of the quark masses differing by more than a factor 3. (This is a reminder of the nearly identical splittings in the charmonium and bottomonium families emphasized above: those are also different  by a similar ratio of 3 in absolute mass scale.)

More specifically, the splittings for the s-shell $ccc$ baryons are calculated to be \be  
M_{ccc}^{2S}-M_{ccc}^{1S}\approx 391, \,\,M_{ccc}^{3S}-M_{ccc}^{1S}\approx  685 \, MeV \nonumber \ee
and for the $uuu=\Delta^{++}$ states they are
\be  
M_{uuu}^{2S}-M_{uuu}^{1S}\approx 438, \,\,M_{uuu}^{3S}-M_{uuu}^{1S}\approx 804 \, MeV \nonumber \ee
The splittings for $uuu=\Delta(3/2^+)$ should be compared to experimental values of these splittings
\be M_{uuu}^{2S}-M_{uuu}^{1S}\approx 278, \,\, M_{uuu}^{3S}-M_{uuu}^{1S}\approx 688 \, MeV, \nonumber
\ee
(from the particle data tables 2022, using  masses from the Breit-Wigner fits). While the deviations between the calculated and experimental values
 are comparable to  the
shifts expected from the spin-spin and 't Hooft forces, which are so far ignored in the calculation, its sign would be opposite, shifting 1S down more than higher states.
(Also, these are relativistic corrections in $\sim 1/M^2$, and would of course be much less important for $c$ quarks.) We therefore
conclude that the effective potential for baryons used (Ansatz A), is somewhat stronger than it is  in reality. We will elaborate on this point in the next section. 

\begin{figure}[h]
\begin{center}
\includegraphics[width=8cm]{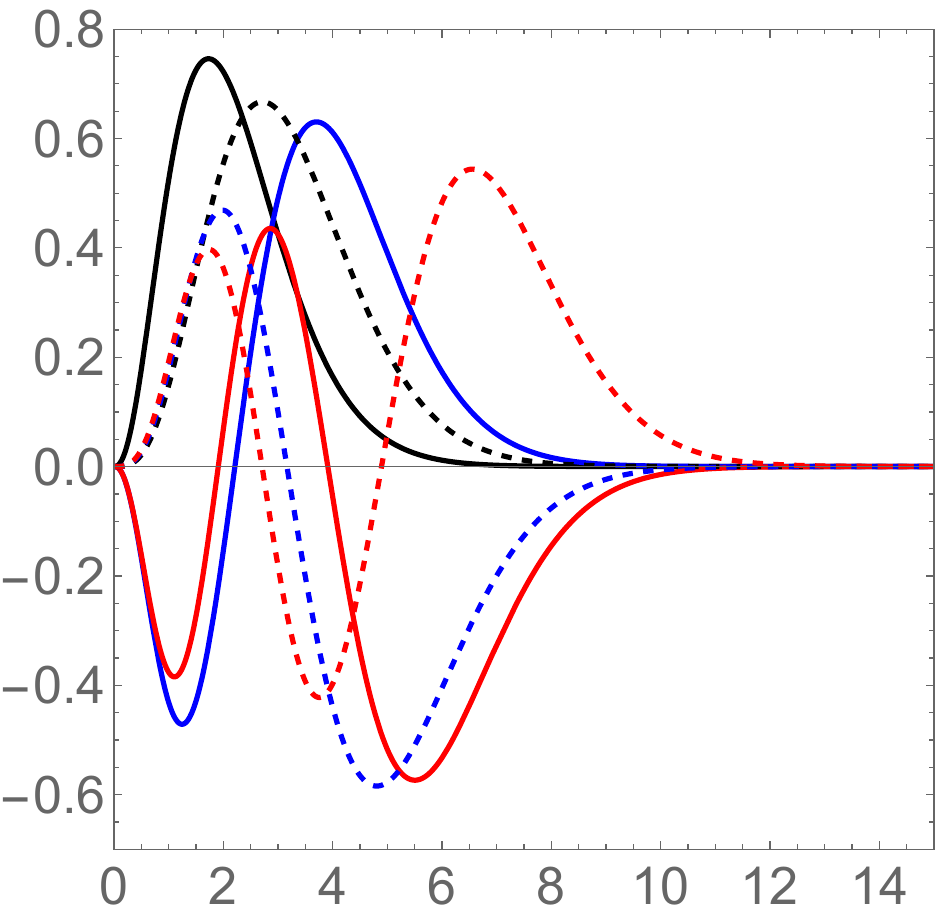}
\includegraphics[width=8cm]{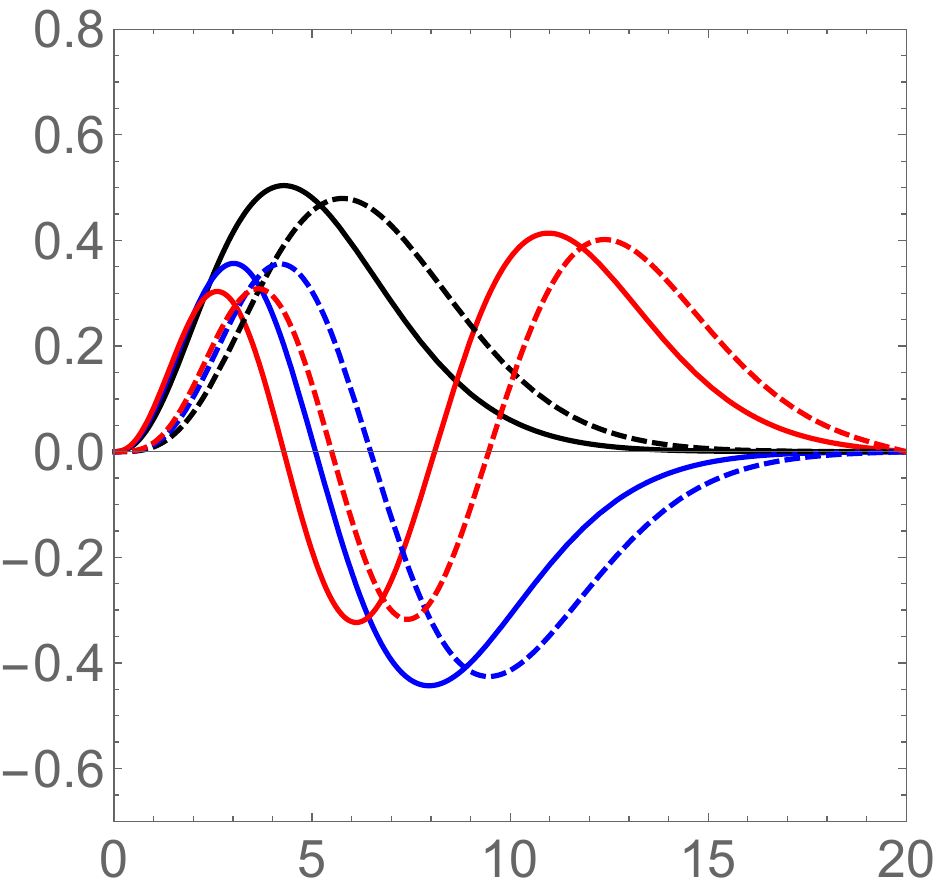}
\caption{Wave functions $u(R_6)$ of $ccc$ (upper) and $uuu$ baryons (lower)  for $L=0$ (solid lines) and $L=1$ (dashed lines), versus the hyper-radial distance $R_6\, (GeV^{-1})$.}
\label{fig_ccc_wfs}
\end{center}
\end{figure}

We also show the corresponding wave functions in Fig.\ref{fig_ccc_wfs}. Unlike the splittings,
those show radical difference between the $uuu$ and $ccc$ baryons. Indeed, the  corresponding hadrons are of quite different sizes, with the $ccc$ ones  much smaller. Also, all of them
 are somewhat unusual for the ground states.
In spite of the enhanced Coulomb forces in the 6d systems, the quasi-centrifugal potential leads to wave functions very strongly suppressed at small hyper-radial distances.

These wave functions can be used for getting matrix elements of various operators. R.m.s radii of three s-states of $\Delta$ are
\be \sqrt{\langle R_6^2 \rangle}=4.97,\,7.53,\,9.72\, GeV^{-1} \ee

To summarize this section: we have tested a popular assumption, that the non-perturbative forces between quarks in baryons 
can be described by a sum of binary forces reduced by 
the same factor 1/2 (as perturbative ones), as compared to mesonic $\bar Q Q$ potential (the so called Ansatz A). We found that it leads to splittings between s-shell states that are somewhat
larger than observed in the $uuu=\Delta^{++}$ baryons. A possible fix is to introduce
an arbitrary factor f for the confining force, and then fit it  experimentally to the splittings if $f< 1/2$.

\subsection{Instanton-induced potentials for baryons} \label{sec_inst_baryons}

We now derive the three-quark effective potential {\em along the radial 6d coordinate} $V_3(R_6)$, using the instanton model of the QCD vacuum.

We start by briefly recalling the approach we took in Ref.\cite{Shuryak:2021hng}. If
no assumptions about the colors of the three quarks are made, the color indices of $W$ 
remain open. Furthermore,
to  model a color-isotropic  vacuum in which instantons
are at random 2-d planes in the $SU(3)$ color space, we introduced
  random  matrices $U\in SU(3)$ which rotate the  instanton fields from their standard $SU(2)$ plane,  $W\rightarrow U W U^\dagger$. 
 All six matrices $U$   were then
averaged  using the invariant {\em Haar measure} of the SU(3)  group,  
\bea
\label{3WP1}
\int dU (U^{a_1}_{i_1} {\bf W}^{i_1}_{1j_1} U^{\dagger j_1}_{b_1})(U^{a_2}_{i_2} {\bf W}^{i_2}_{2j_2} U^{\dagger j_2}_{b_2})(U^{a_3}_{i_3} {\bf W}^{i_3}_{3j_3} U^{\dagger j_3}_{b_3})\nonumber\\
\eea
before convolution of the external color indices. In~\cite{Shuryak:2021hng} we used
the so called 6-$U$ ``Weingarten formula" for this integral, and the Wilson lines running through instanton fields, leading to a  ``generic triple-quark
 potential" of the form
\bea
\label{3WP3}
V&=&n_{I+\bar I}\int d^3z
\bigg[(1-c_1c_2c_3) \delta^{a_1}_{b_1}\delta^{a_2}_{b_2}\delta^{a_3}_{b_3}  \nonumber \\
&+&\frac{9}{8}\,c_1 s_2 s_3 n_2\cdot n_3 \delta^{a_1}_{b_1}\bigg(\frac 12 \lambda_2^B\bigg)^{a_2}_{b_2}\bigg(\frac 12 \lambda_3^B\bigg)^{a_3}_{b_3} \nonumber \\
&+&{\rm 2\, perm.}\bigg] 
\eea
with  trigonometric functions involving  color rotation angles, that depend on the 3-dimensional distances $\vec \gamma_i,i=1,2,3$ between the 
location of the Wilson lines $\vec r_i$, and  the instanton center $\vec y$
\bea
\label{LOC}
c_i&\equiv&{\rm cos}(\alpha_i), \,\,
s_i\equiv {\rm sin}(\alpha_i)\nonumber\\
\gamma^2_i&=& 
(\vec r_i-\vec y)^2, \,\,\, \vec n_i=\vec \gamma_i/ |\vec \gamma_i |
\eea
Since the standard instanton is an $SU(2)$ solution, it does not interact with a quark with the ``third" color, orthogonal to the $SU(2)$ plane. The third
color was ignored in the formula above. (Below we will extend the $2\times 2$
matrix by adding $W_A(3,3)=1$ and other zeros  to the $3\times 3$ matrix.) Yet, for random orientations of the instanton induced by a rotation matrix $U$, 
and  for arbitrary colors of the three quarks, the potential is generically a three-body potential, $not$ a sum of two-body ones.

Here, we perform a different calculation,
limiting it   to the case of three quarks {\em making  baryons}. The corresponding color wave functions are  $C_{ijk}= \epsilon_{ijk}/\sqrt{6}$. Putting those before and after the Wilson lines with the instanton-induced rotations (\ref{3WP1}),  the result 
simplifies considerably. The randomizing matrices $U$ acting on the Levi-Civita tensor cancel out (unit determinant). The resulting potential is therefore reduced to what we will call a ``determinantal form" 
\ba \label{eqn_V_3}
V_3 &=& (n_{I+\bar I}\rho^3)\int {d^3 y \over \rho^3}\\
&\times & \bigg( 1- \sum  {\epsilon_{a_1 a_2 a_3} \over \sqrt{6}} W_1^{a_1 b_1} W_2^{a_2 b_2} W_3^{a_3 b_3} {\epsilon_{b_1 b_2 b_3}\over \sqrt{6}}\bigg) \nonumber
\ea
Note that far from the instanton the color rotation angles of the Wilson lines (\ref{eqn_angles}) vanish. All $W_{ab}\rightarrow \delta_{ab}$ and therefore the integrand goes to zero, with a converging integral. Note also that we rewrote
the integral over the instanton location $d^3y$ as dimensionless, with a coefficient 
$(n_{I+\bar I}\rho^3)\approx 51\, MeV$. 

Although this expression has the form of a generic three-body interaction of three Wilson lines, for Wilson lines in the field of an  instanton,  we (with some surprise) found that it does become the sum of three $binary$ interactions for 12,13,23 pairs, each proportional to the ``relative rotation"
\be 
cos(\alpha_i)cos(\alpha_j)+sin(\alpha_i)sin(\alpha_j)(\vec n_i\cdot \vec n_j) 
\ee
The same combination appears in the $\bar Q Q$ potentials.

In Ref.\cite{Shuryak:2021hng} we evaluated the  potential for different triangles,
and compared the results to similar triangles studied on the lattice.
Here we analyse  baryons in the hyperspherical approximation, so
we just take one equilateral triangle with variable size. The quarks are put at the locations
 $(r,0,0),(0,r,0),(0,0,r)$, so the distances between them are $r_{ij}=\sqrt{2}r$.
 Using the definitions of the Jacobi coordinates, one can see that $R_6=\sqrt{2}r$
 as well. 

\begin{figure}[h]
\begin{center}
\includegraphics[width=7cm]{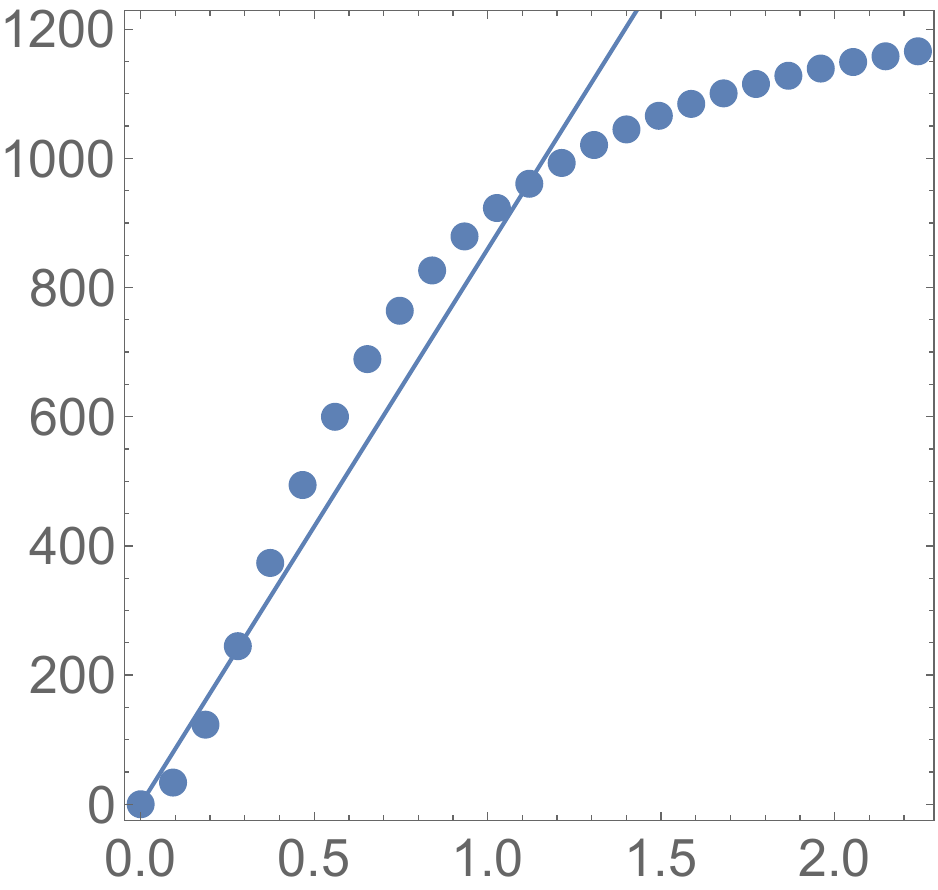}
\caption{The points show the instanton-induced effective potential in baryons $V(R_6), (MeV)$
vs $R_6\, (fm). $. The tension force is of the order of
 1 GeV/fm (shown by a line for comparison).}
\label{fig_WWW_pot}
\end{center}
\end{figure}

The effective three-quark potential $V_3$ in (\ref{eqn_V_3}) for this equilateral triangle,  is shown in Fig.~\ref{fig_WWW_pot}.
Here we used $n_{I+\bar I}=7\, {\rm fm}^{-4}, \rho=1/3\, {\rm fm}$, corresponding to the ``dense instanton vacuum model" we used for the meson potential.  Remarkably,
the effective $QQQ$ potential along $R_6$, is quite similar in shape to the meson
$Q\bar Q$ potential.

Using this potential (with the same one-gluon exchange as in previous section) in a
 hyperspherical approximation, gives the following splittings  in the $ccc$ and $uuu$ baryons 
 (GeV)
 \ba M_{ccc}^{2S}-M_{ccc}^{1S}&=& 450,   \,\,\,  M_{ccc}^{3S}-M_{ccc}^{1S}=675 , MeV \nonumber \\
 M_{uuu}^{2S}-M_{uuu}^{1S}&=&  232, \,\,\,  M_{uuu}^{3S}-M_{uuu}^{1S}= 394\, MeV  \nonumber
 \ea
 While the results for the $ccc$ baryons are very close to those found in the previous section
 for ``ansatz A",  those for the $uuu$ baryons are reduced significantly, and are now much closer to the experimental value of the first splitting. This means that the
 instanton-induced potential is a better representation of the force at $R_6\sim 1\, fm$.
 The splitting of the 3S state is too small because the instanton-induced
 potential at $R_6 > 1 {\rm fm} $ is flat. In order to get it in agreement
 with the data one would need a potential in between linear and constant,
 perhaps growing as a slower function of the distance.




\section{The all-charm tetraquarks }
\subsection{Masses and models} \label{sec_tetra_intro}
Experimentally these type of tetraquarks have been discovered by LHCb \cite{2006.16957} and then
by CMS \cite{2212.00504} and ATLAS \cite{2209.12173} in $J/\psi,J/\psi$ and related channels. These states are named X(6550), X(6900), X(7280).
For definiteness, we will use the ATLAS masses
 \be M_1=6620,\,\,\, M2 = 6870,\,\, M3 = 7220 \, (MeV)\ee
 all with errors $\pm 30 \, MeV$.  These states
 provide a good opportunity to check our understanding
 of both the multi-particle dynamics and inter-quark forces.

Among the many models put forth for the tetraquarks, we note the diquark-antidiquark formulation~\cite{2009.04429,2008.01631}, in which a 4-quark system is represented as a binary state of diquarks with binary potentials, 
equal to that in quarkonium $\bar Q Q$. This would be true
if distances between quarks in diquarks (to be called $r$ below)
are much smaller than  distances between diquarks $R$.

Since they are supposed to be comparable for $r\sim R$
in the lowest tetraquark states, we cannot  neglect any of the 6 pair-wise interactions in the
four-body compounds. 
 In this section we  use the four-body approach, with all six binary potentials included. 
%
Lattice studies of four-quark static potentials, with quenched and dynamical quarks, have been performed in \cite{1702.07789}. (Again, the dependence on the hyper-distance could be determined, but was not.)

\subsection{The lowest all-charm tetraquark states in the  hyper-distance approximation} \label{sec_4body_Sch}

The use of the hyper-distance approximation for all-charm teraquarks was pioneered in \cite{Badalian:1985es} (see also the recent study in~\cite{Badalian:2023krq}). 
The  main idea is that
for the lowest states, one may assume that  they are mostly
$L=0$ s-shell, with only a small d-shell admixture.
If so, we only need to solve a one-dimensional radial Schroedinger equation.

For four particles, there are three Jacobi coordinates, or 9-dimensions. We note that in our definition of these coordinates and hyper-distance (see Appendix \ref{sec_kinematics_cccc}), as well as the ensuing 
radial equation, we differ with~\cite{Badalian:2023krq}.
Also, our  9-dimensional wave function is written 
with a different power \be \psi(R_9)=u(R_9)/R_9^4 \ee
which allows the elimination of the
first derivative term in the 9-d Laplacian, and also puts the
 normalization into the 1-dimensional  form  $\sim \int dR_9 | u(R_9) |^2$. The corresponding radial Hamiltonian
 is then
\be \label{eqn_hyper_Sch}
\bigg(-\frac{d^2}{dR_9^2}+ {12 \over R_9^2}\bigg){1 \over 2M}+(4M+V)
\ee
in which the second term in the bracket is another ``quasi-centrifugal" term, following from the 9-dimensional  Laplacian. 

The potential includes the color factors  derived in Appendix~\ref{sec_kinematics_cccc}, which happen to be the same for $\bar 3 3$
and $6\bar 6$ diquark color structures.
The other factors stemming from the angular projection
of the interparticle distances $r_{ij}$ to the hyper-distance $R_9$,
are discussed in Appendix~\ref{sec_kinematics_cccc} for the Cornell-type
potentials. Here we use the Cornell potential as used for the Upsilons above, 
with
\be 
V = 2*(0.773*R_9*0.113 - 1.55*0.64/R_9) 
\ee

As for the mesons and baryons we discussed above, and
in order not to get bogged down by the issue of the quark mass value and the
constant in the potentials, we focus  on the level $splittings$.
Solving the Schroedinger equation (\ref{eqn_hyper_Sch}),  we find for the of the s-shell splittings 
\ba M(2S)-M(1S)&\approx& 370\, MeV \nonumber \\
M(3S)-M(1S)&\approx& 689\, MeV \nonumber \\
M(4S)-M(1S)&\approx& 975 \, MeV
\ea
and for the p-shell splittings
\ba M(1P)-M(1S)&\approx& 587\, MeV \nonumber \\
M(2P)-M(1S)&\approx& \,879\, MeV \ea 
although the latters are not good candidate states
for decaying into $J/\psi$ pair.

From experimentallly reported  (ATLAS fit) masses of the three $cc\bar c \bar c$ resonances, we find for the splittings
\ba \label{eqn_atlas_splittings}
M(2)-M(1)&\approx& 250\, MeV \\
M(3)-M(1)&\approx& 600\, MeV \nonumber \ea
They are in fair agreement with the  S-shell levels we calculated.
As experience shows, this means that the overall magnitude of the forces in all the 6 quark pairs, is in reasonable agreement with the reported observations.

\begin{figure}[h]
\begin{center}
\includegraphics[width=8cm]{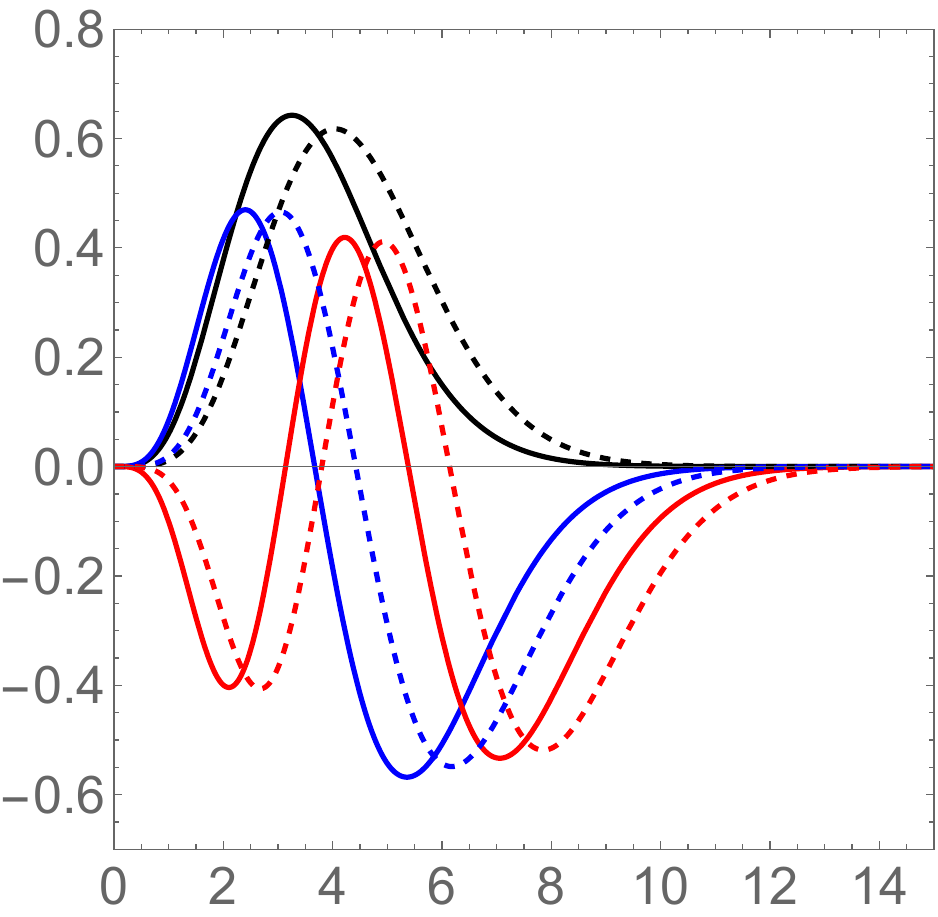}
\caption{ The wave functions $u(\rho)$ for 1S,2S,3S states of all-charm tetraquarks in hyperspherical approximation, shown by solid black,blue and red lines, respectively. The dahsed lines are p-shell states.}
\label{fig_tetra_wfs}
\end{center}
\end{figure}

The corresponding wave functions are shown in Fig.~\ref{fig_tetra_wfs}. We note again, the  very strong suppression at small hyper-distance, which is due to the quasi-centrifugal term in the Laplacian. (In \cite{Badalian:2023krq} it is noted that
it coincides with the centrifugal term for the $L=3$ or F-shell quarkonia.)
The r.m.s. sizes of these states are therefore rather large in
$\, GeV^{-1}$, respectively. 

 \section{Instanton-induced forces in tetraquarks} \label{sec_tera_inst}

The problem with the correlators of four Wilson lines, like for three ones,  can either
be set for (i) arbitrary colors of the quarks, or (ii) for those
in a particular color wave function. In the former case,
the correlator has color indices open and each Wilson line is
rotated by random matrices $U W U^\dagger$, with subsequent averaging
over the $U$ using the invariant measure of the $SU(3)$ group. The general case
of eight $U$  is not used,  but still  discussed in Appendix~\ref{sec_Weingarten}.

\begin{figure}[h]
\begin{center}
\includegraphics[width=7cm]{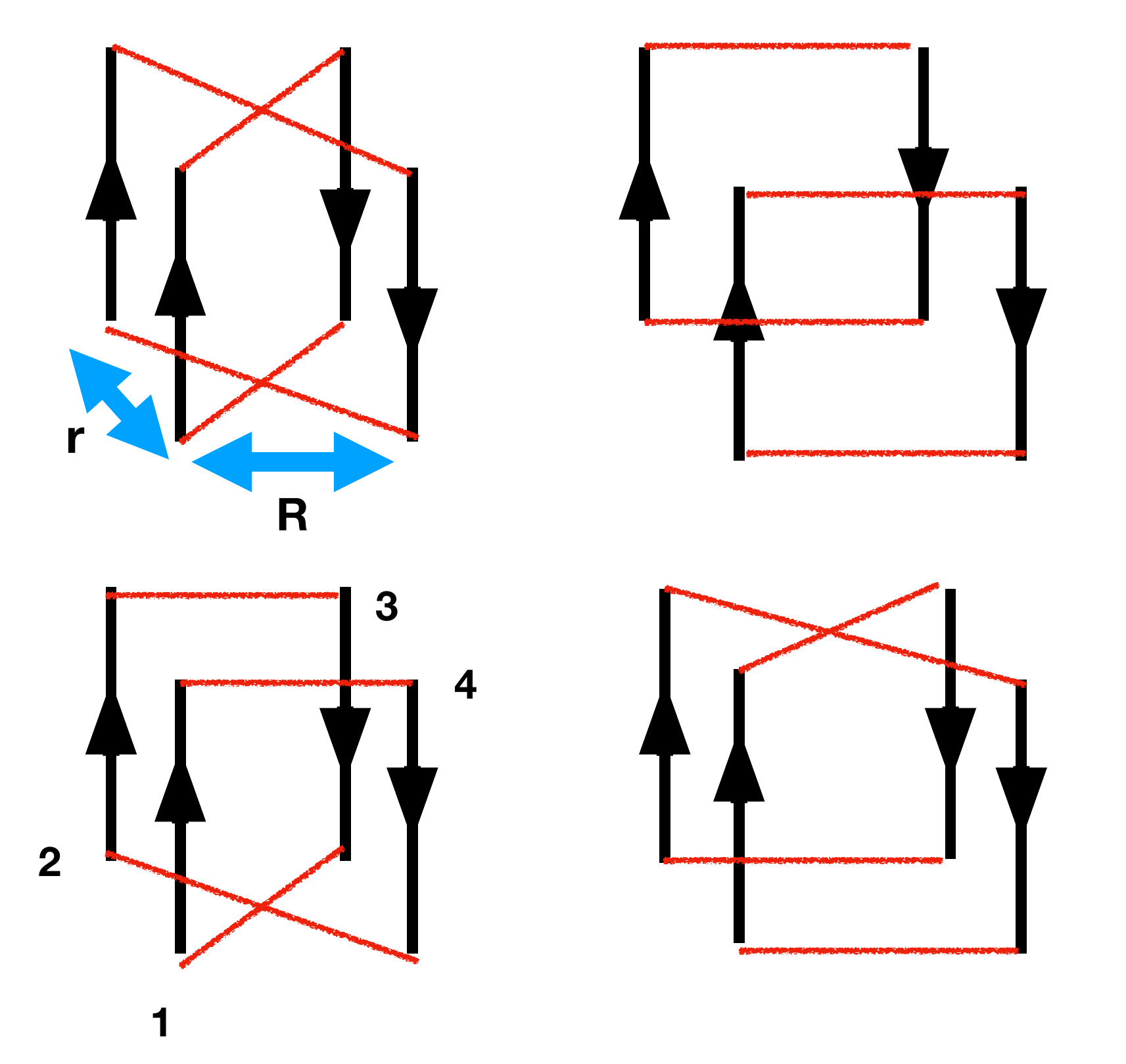}
\caption{The correlators of four Wilson lines in tetraquarks, connected
according to the product of color wave-functions (\ref{eqn_diquark_wfs})
at times $\pm T$. The arrow directions show quarks (1 and 2) and antiquarks (3,4). In the $\bar 3 3$ case the 
convolutions in the second row appears with the minus sign, and in the 
$ 6 \bar 6$ case all terms have plus sign.}
\label{fig_WWWW}
\end{center}
\end{figure}

The color wave functions built from diquarks are discussed in  Appendix~\ref{sec_kinematics_cccc}. Applying those at early and late times, we get certain products of delta indices. The convolution of the color indices  is shown by red lines in Fig.\ref{fig_WWWW}.  The analytic expression for the $\bar 3 3$ correlator is
\ba \label{eqn_K}
&&K_{\bar 3 3}(r,R)= \\
&&\bigg< \sum C_{\bar 3 3}^{c'_1,c'_2,c'_3,c'4} W_{c'_1}^{c_1} W_{c'_2}^{c_2} W^{c'_3}_{c_3} W^{c'_4}_{c_4} C_{\bar 3 3,c_1,c_2,c_3,c_4}\bigg>_y \nonumber
\ea
where the color wave functions are given in (\ref{eqn_diquark_wfs}). The sum is assumed over all color indices, and the
averaging corresponds to the 3-dimensional integration over the instanton center position $\vec y$. Note that the anti-quark $W$ (assumed here to be 3,4) are shown as transposed, yet by including the opposite
color charges this is equivalent to Hermitian transposition of the corresponding
unitary matrices. In effect, the correlator corresponds to matrix products in the order indicated in Fig.~\ref{fig_WWWW}.
The same expression with the sextet wave-function is used for the $6 \bar 6$ correlator.

Although the diquarks on which the classification is based are pairs of $qq$
or $\bar q \bar q$, the delta functions connectors are between quarks and
antiquarks, like in mesonic case. Therefore,  
the resulting structures consists of  product of two traces (upper row) or a single trace (lower row) of Wilson lines products. Like in the baryon case, these traces do not change if
$W$ are all rotated by some matrices $U$. Again there is 
no need to use $SU(3)$ averaging of $U$ (via Weingarten-style formulae).

\begin{figure}[h]
\begin{center}
\includegraphics[width=7cm]{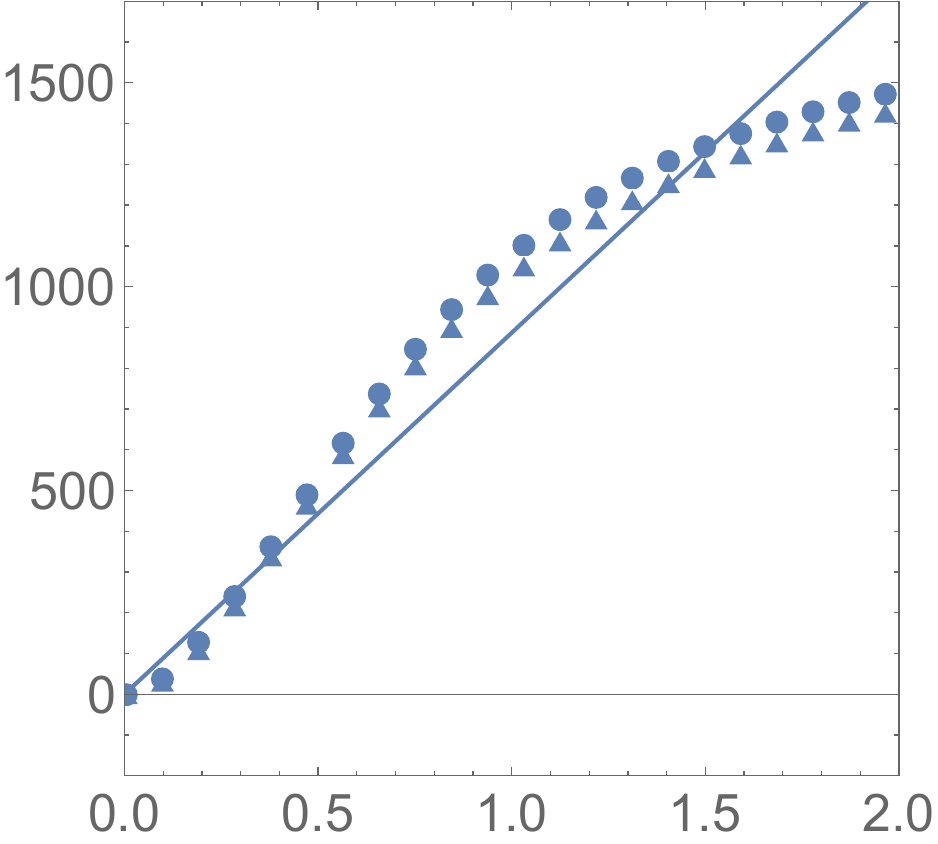}
\caption{Four-quark effective static potentials (in MeV) versus the hyperdistance $R_9$ (in fm), calculated from the correlators of four Wilson lines in tetraquarks. The color indices are connected
according to products of color wave functions at $T\rightarrow \pm \infty$(\ref{eqn_diquark_wfs})
at times $\pm T\rightarrow \pm\infty$.  The results for the
$\bar 3 3$ channel
are shown by triangles, and that for the $6\bar 6$ 
channel by closed points.
The line corresponds to the non-perturbative part of the
Cornell potential projected to the hyper-distance $R_9$, as described in Appendix~\ref{sec_jacobi_baryons}.}
\label{fig_66_33_vs_R9}
\end{center}
\end{figure}

While four $W$ can be located arbitrarily, 
we have only calculated the instanton contributions to  correlators of four Wilson lines,  placed at the corners of a $r\times R$ rectangle. 
The quarks are set to be 1,2 at distance $r$ from each other, and the distance between the diquark and the antidiquark is set to be
 $R$. The potential, as usual, is defined as
\be 
V_{\bar 3 3}(r,R)=n_{I+\bar I}\int d^3 y \bigg(1- K_{\bar 3 3}(r,R)\bigg)
\ee
where, we recall that the calculation is basically the 3-dimensional integration over the location of the instanton center $\vec y$. 
For the $6\bar 6$ case, the same formula applies,  with the pertinent color 
wave-function. In Fig.~\ref{fig_66_33_vs_R9} we show only the 
results for the potential,
from the symmetric ``square setting" of charges, $r=R= R_9/\sqrt{2}$, its dependence
on the overall size of the tetraquark, or the hyper-distance
in 9 dimensions.

A key question is whether the $\bar 3 3$  and 
 $ 6 \bar 6$ channels, support the same effective radial potentials. 
Our results in Fig.\ref{fig_66_33_vs_R9}, show that $6\bar 6$  and $\bar 3 3$ potentials are not identical.
An inherent repulsion in the  $ 6 \bar 6$ channel
generates an extra increase in the potential, especially 
at large $R_9> 1\, fm$. 
We conclude that the instanton-based forces {\em approximately} support equality of the potentials in both channels, as anticipated in~\cite{Badalian:1985es}. 
 
As in all similar calculations at small $R_9\rightarrow 0$, the strong cancellations between the quark and antiquark contributions,
lead to a vanishing effective potential .
The overall shape of the potential is similar (but not equal) to 
other instanton-induced potentials (e.g. $qqq$ one in Fig.\ref{fig_WWW_pot}).

At large distances, the potentials tend to  constants,
since the instanton vacuum does not confine. However, the approach
to these constants is from below, and does not happen even at
$R_9=2\, fm$, the maximal value  shown in the figure. 

Finally, we show by a straight line the non-perturbative part
of the Cornell potential used in the previous section. The overall force (potential gradient) is  quite
close to the instanton-induced force.

Using the instanton-induced potentials into the radial Schroedinger equation in the $\bar 3 3$ channel, 
we have for  the s-level splittings
\ba \label{eqn_inst_tetra_splittings}
M(2)-M(1)&\approx& 362\, MeV \\
M(3)-M(1)&\approx& 578\, MeV \nonumber \\
M(4)-M(1)&\approx& 657\, MeV \nonumber \ea
The comparison to the empirical splittings in  (\ref{eqn_atlas_splittings}), shows satisfactory  agreement.
We emphasize that the
the calculation of the radial potential does not involve
any free parameters, as we made use of  the same instanton ensemble 
for also the $\bar q q$ and $qqq$ channels.

\section{Summary and discussion} \label{sec_summary}
We conclude, by first recalling some of the  questions of hadronic spectroscopy we have tried to address. A central one is the nature of the effective static potential for quarkonia, including
the perturbative contribution one-gluon-exchange at short distance,  and the non-perturbative and about  linear contribution at large distance. It has approximately the same tension
in quarkonia as deduced from Regge trajectories of light mesons and baryons. 

In this paper we have  addressed  the effective interaction potentials in two types of few-body hadrons --
 baryons and tetraquarks.
While for the one-gluon-exchange modifications by  color factors can  be  easily calculated
 (resulting e.g. in the known factor  $\frac 12$  in baryons),
 the corresponding factors in the non-perturbative part is quite nontrivial. 
 It is even unclear whether it can or cannot be dominated by a sum of 
 binary interactions.
 
 For simplicity and symmetry, 
  we usde the so called ``hyperdistance approximation", projecting the relevant multidimensional potentials to a single radial variable $R_6$ or $R_9$,
  in the space of Jacobi coordinates.
The squared hyper-distances are proportional to the sum of the squares of all the binary distances,  in a very symmetric way.
Solving radial Schrodinger equation one finds spectra and wave functions, in 6 and 9 dimensions respectively. We focused on the splittings between the subsequent s-shell levels, known to be rather insensitive to the quark masses, but roughly proportional to the force between the constituents. We found that the Cornell-type
potential, averaged over solid angles in these spaces and with proper color factors, does indeed generate reasonable values for $uuu=\Delta^{++}$ baryons, 
and for the recently discovered all-charm tetraquarks.

The static spin-independent potentials are known to be related to correlators of few (3,4 etc) Wilson lines, connected in ways depending on the external color wave functions of the given hadron.  
 Such objects are very non-local correlators
of vacuum fields, with a non-trivial dependence on the locations of the quarks.
{\em A priori}  it not possible to use a  traditional notion of the potential energy being a sum of 
 binary potentials. For corresponding lattice studies,
  see e.g. \cite{Koma:2017hcm} for baryons and \cite{1702.07789} for tetraquarks.

 The Euclidean gauge fields in the QCD vacuum
can be, in a semiclassical approximation, approximated by $instantons$, 4-dimensional solitons (pseudo-particles) describing tunneling events through
topological barriers which separate configurations with different Chern-Simons numbers. Models of their ensembles were developed, both using phenomenology of hadronic correlation functions (for review see e.g.\cite{Schafer:1996wv}) and direct
lattice studies of the fields. The instanton fields are special because they are giving
 simple analytic answers for Wilson lines, and therefore one can derive rather simple
analytic expressions for their correlators~\cite{Callan:1978ye}. Basically, the Wilson lines
are certain rotation  in the quark color spaces, by an angle
that depends  on the distance to the instanton center.

In \cite{Shuryak:2021fsu} we introduced and used the  ``dense instanton liquid model" to evaluate static and spin-dependent potentials in
quarkonia. In this paper we continue this work, by evaluating the instanton-induced potentials via
correlators of Wilson lines, for baryons and tetraquarks.

The instanton-induced potential for baryons is derived in section 
\ref{sec_inst_baryons}. Somewhat surprisingly, we found 
that the triple-W correlators  can be written as 
 a sum of binary potentials. Furthermore, its  shape
happens to be the same as for quarkonia, differing just by an overall factor. We think that the
main reason for this,  is the fact that the instanton
fields are always inside a certain $SU(2)$ subspace of the $SU(3)$
color group, while the color wave functions of baryons require that the three quarks  carry always  three mutually orthogonal colors. 

We have evaluated the instanton-induced potential for tetraquarks.
In this case the color wave functions are different. 
We have considerec two options, which we referred to 
as $\bar 3 3$ and $6\bar 6$ according to their diquark color content.
The convolution  with four Wilson lines leads to certain diagrams shown in Fig.\ref{fig_WWWW}. Apparently in this case,  no splitting of the 4-body interactions into pairwise
binary terms takes place. Yet the effective potentials
for the $\bar 3 3$ and $6\bar 6$ channels as a function of the hyperdistance $R_9$, shown in 
Fig.\ref{fig_66_33_vs_R9}, are similar in shape to those in quarkonia and  $qqq$ cases.

Before we made use of these potentials in spectroscopy, 
in section \ref{sec_inst_meson} 
we returned to the  (rather well-studied) case of heavy quarkonia,
and in particular to the relation between their spectra and the effective potentials.
In order not to deal with the subtleties related to the quark masses and additive terms in the potentials, we focused on the $splittings$ between the s-shell levels, using the
standard 3-dimensional Schroedinger equation. We have seen that while the Cornell-type
linear+Coulomb potentials do a very good job, our instanton-induced 
potential can also be used, with a precision even better than the fitted Martin potential. 

Proceeding to the baryons and tetraquarks,  we used a very similar approach, based
 on the so called ``hyper-spherical approximation", or ``method of K-harmonics" known since 1960's. It is based on the assumption 
 that the lowest few-body states are ``hyper-spherically symmetric", in the
 6-dimensional space for baryons, and 9-dimensional
space for tetraquarks, in terms of pertinent Jacobi coordinates. 
Solving the radial hyper-spherical Schroedinger equation,  we obtained the spectra,
in particular the splittings between s-shell levels. For baryons we did so
 for the $ccc$  and $uuu$
flavor-symmetric baryons. While the formers are not yet experimentally observed, the latters belong to the family of $\Delta^{++}$ for which
we found a reasonable agreement between the derived effective potential, and the experimentally observed separations of the (s-shell) states.

In section \ref{sec_4body_Sch}, following the pioneering paper \cite{Badalian:1985es}, we applied the
 hyper-spherical approximation to  all-charm tetraquarks. This
method assumes that lowest the states are ``hyper-spherical" in 9-dimensional
space of Jacobi coordinates. The interaction assumed to be the sum of 6 binary potentials of the Cornell type times $\lambda_i^A \lambda_j^A$
color matrices, times some extra effective strength factor. Focusing  not
on masses themselves, but on the $splittings$ between the states, we simplified
the problem by eliminating the overall constants, a conspiracy between quark masses and  constants in the potentials. Our instanton-based potentials in $\bar 3 3$
or $6 \bar 6$ channels are very close, so in a way we confirmed the conjecture in~\cite{Badalian:1985es} that they are the same, in spite of certain repulsive terms in the $6 \bar 6$ case.
Furthermore, this potential leads to $1S-2S-3S$  splittings (\ref{eqn_inst_tetra_splittings}) in fair
agreement with the observed splittings between the three recently reported resonances  (\ref{eqn_atlas_splittings}). We also compared our results to the recent analysis in
\cite{Badalian:2023krq}, based on the hyper-distance approximation as well.

Finally, we stress once again that the 
hyper-radial $qqq$ and $cc \bar c \bar c$
effective potentials we derived, Figs.\ref{fig_WWW_pot} and \ref{fig_66_33_vs_R9},  originate without change from the instanton model of the QCD vacuum. Its two parameters are fixed by  lattice 
observation of instanton, and therefore one may say that the present paper has
$no$ free parameters whatsoever. The hadronic masses and wave functions we derived, follow directly from a model of the QCD vacuum.
\\
\\
\\
{\bf Acknowledgements:}
\\
This work  is supported by the Office of Science, U.S. Department of Energy under Contract No. DE-FG-88ER40388.

\appendix

\section{Hyper-spherical harmonics} \label{sec_hyper_general}
The non-relativistic kinetic energy of $A$ particles using $N$ Jacobi coordinates, takes the simple form~\cite{FABREDELARIPELLE1979185}
\be
-\frac 12 \sum_{i=1}^A\,\nabla^2_{i}=
-\frac 12 \bigg(\partial_\rho^2+\frac{3N-1}\rho\,\partial_\rho-\frac 1{\rho^2}K_N^2\bigg)
\ee
The hyper-spherical harmonics (HHs) are the eigenstates of the grand-angular momentum 
\be
K_N^2\,{\cal Y}_{[K]}^{KLM_L}(\Omega_N)=(K(K+3N-2))\,{\cal Y}_{[K]}^{KLM_L}(\Omega_N)
\ee
The $2^N+N$ angles $\Omega_N=(\hat{\xi}_1,..,\hat{\xi}_N ; \varphi_1, .., \varphi_N)$,
are those of the individual Jacobi coordinates $\hat{\xi}_i$, and the hyperangles
${\rm cos}\varphi_j=\xi_j/\rho$. 
$L,M_L$ are the standard 
quantum numbers of the total orbital angular momentum $L^2, L_z$.
The specific form of the HHs follows by recoupling the individual angular momenta $l_i$.  They are normalized as
\be
\int d\Omega_N\,{\cal Y}_{[K]}^{KLM_L\ *}(\Omega_N)\,{\cal Y}_{[K']}^{K'L'M'_L}(\Omega_N)
=\delta_{[K],[K']}
\ee
and their total number  is~\cite{FABREDELARIPELLE1979185}
\be
d_K=(2K+3N-2)\frac{(K+3N-3)!}{K!(3N-2)!}
\ee
For instance for $A=4$ particles with $N=3$, 
the $K=0$ HH has degeneracy $d_0=1$, and the $K=1$ HHs have degeneracy $d_1=9$.

\section{Quantum mechanics of three quarks in hyperspherical coordinates} \label{sec_jacobi_baryons}
To describe the specifics of the hyper-spherical approximation, 
we first consider three quarks in a baryon. 
 The three coordinate vectors of the quarks are compressed into two (modified) Jacobi vectors \bea 
\vec\rho=\frac 1{\sqrt{2}} \vec r_{12}\qquad \vec \lambda=\frac{1}{\sqrt{6}}(\vec r_{13}+\vec r_{23})
\eea
with $\vec r_{ij}=\vec r_i-\vec r_j$ as 3-dimensional  vectors.
The 6-dimensional radial coordinate denoted by $R_6$ is given by
\be R_6^2=\vec\rho^2+\vec \lambda^2={1\over 3} [(\vec r_1-\vec r_2)^2+(\vec r_1-\vec r_3)^2+(\vec r_3-\vec r_2)^2]\ee
so it is related to the sum of all the squared interquark distances, in a symmetric manner. The coefficients in the definitions of these coordinates also
ensure that the kinetic energy takes the form of a Laplacian in 6-dimensional space  $\{\vec \rho,\vec \lambda \}$, divided by $2m$. (The total momentum  has a
different coefficient, but it is presumed to be zero anyway.)

The hyper-spherical approximation in its most symmetric form does not distinguish
$\vec \rho,\vec \lambda$, but treats all 6 coordinates on equal footing. Standard spherical coordinates can be introduced with 5 angles $\theta_1,\theta_2,\theta_3,\theta_4\in [0,\pi],\phi\in [0,2\pi]$, and a 6-dimensional  solid angle 
\ba \Omega_6 &=& \int d\theta_1 d\theta_2 d\theta_3 d\theta_4 d\phi\\
 & \times & sin(\theta_1)^4 sin(\theta_2)^3 sin(\theta_3)^2 
  sin(\theta_4)=\pi^3 \nonumber \ea
(Alternative a 6-dimensional coordinate space parametrization with
 five angular variables can be defined as $\vec \rho=R_6 cos(\Phi) \hat\vec n_1,\vec \lambda=R_6 sin(\Phi) \hat\vec n_2$ with two unit vectors
 having independent solid angle integrals.) 

The effective potential $V_6$ in general depends on all 6 coordinates, but can
be ``projected 
  onto the 6-dimensional radial coordinate $R_6$" by angular average. If the binary potentials are
  of Cornell-type, a combination of linear and inverse terms, e.g.
$V_{12}=\kappa_2/r_{12}+\sigma_2 r_{12}$, then 
their projections are done via the solid angle integral
\ba \langle r_{12} \rangle &=& \sqrt{2}\langle |\vec \rho | \rangle  
=\sqrt{2}R_6
{\int d\Omega_6 d_{12} \over \Omega_6} \nonumber \\
&\approx & 0.960 R_6  \ea
where 
\be  d_{12} =\sqrt{C_1^2 + S_1^2C_2^2 + S_1^2S_2^2C_3^2}=\sqrt{1-S_1 S_2 S_3} \ee
with the short-hand notations $C_i=cos(\theta_i), S_i=sin(\theta_i)$.

A similar averaging of the inverse power of $r_{12}$ has $1/\sqrt{2}d_{12}$ in the integral
\bea 
\bigg<{1 \over r_{12}} \bigg> &=&{1 \over \Omega_6 } \int d\Omega_6 {1 \over \sqrt{2} d_{12}}  \nonumber \\
&\approx & 1.20/R_6 
  \eea
Note that the convergence of this integral is due to the fact that $d_{12}$
can only vanish if all three angles $\theta_1,\theta_2,\theta_3\approx \pi/2$.
As a result, the effective potential in hyper-spherical coordinate $R_6$ takes the form
\be \label{eqn_Cornell_6d}
V(R_6)= 3\big[0.96 *R_6*(0.113/2)- 1.2 R_6*(0.64/2)\big]
\ee
where the factor 3 appears due to the inclusion of interactions of $12,13,23$ quark pairs. The   numerical values $0.113,0.64$ 
(in GeV$^2$ units) are from our version of the Cornell potential (\ref{eqn_Cornell}), and the division by two comes from assumption of Ansatz A for the $qq$ relative to the $\bar q q$ potential.

\section{Quantum mechanics of four quarks in hyper-spherical coordinates} \label{sec_hyper_tetra}
The generalized  Jacobi coordinates  for 4 particles are
defined by
\ba  \vec\xi_1 &=& \sqrt{1 \over 2}(\vec r_1-\vec r_2) \\ 
\vec\xi_2 &=& \sqrt{1 \over 6}(\vec r_1+\vec r_2-2\vec r_3) \nonumber \\
\vec\xi_3 &=& {1 \over 2 \sqrt{3}}(\vec r_1+\vec r_2+\vec r_3-3\vec r_4) \nonumber \ea
In this case the hyper-distance 
\ba R_9^2&=&\vec \xi_1^2+\vec \xi_2 ^2+\vec \xi_3^2 \\
&=& {1\over 4} [(\vec r_1-\vec r_2)^2+(\vec r_1-\vec r_3)^2+(\vec r_1-\vec r_4)^2  \nonumber \\
&+&(\vec r_3-\vec r_2)^2+(\vec r_4-\vec r_2)^2+(\vec r_3-\vec r_4)^2]
\nonumber \ea
 is connected to the sum of the squared distances  of all six pairs of quarks. When supplemented  by
 8 angles, it describes the 9-dimensional space in which quantum mechanics is performed. Again, with the total momentum set to zero, the kinetic energy is given by the 9-dimensional Laplacian with the standard coefficient 
\be K= -{1 \over 2 m_c}\Delta_\xi \ee
The corresponding  calculations are in section \ref{sec_4body_Sch}. 

(To our surprise, in \cite{Badalian:2023krq}  $non$-Jacobi coordinates 
are used, namely
 \ba \vec\xi_1 &=& \sqrt{1 \over 2}(\vec r_1-\vec r_2) \\
 \vec\xi_2 &=& {1 \over 2} (\vec r_3+\vec r_4-\vec r_1-\vec r_2) \nonumber \\
 \vec\xi_3 &=&\sqrt{1 \over 2}(\vec r_3-\vec r_4) \nonumber \\
 \ea
for which neither the radial coordinate is the symmetric sum of all distances as above, nor the kinetic energy is the Laplacian. ) 

The dependence on the hyper-distance only means that we deal with the lowest $K=0$ or s-shell.  Only the radial Schroedinger equation needs to be solved. Note that
after changing to the reduced wave-function,
it differs from the familiar 3-dimensional case, by only the
``quasi-centrifugal" term $12/R_9^2$. 

The angular averaging is performed as in the previous section, except now
there is a different volume element, namely
\ba \Omega_9&=&\int (\prod_{i=1}^7 d\theta_i) d\phi \\
&\times & sin(\theta_1)^7*sin(\theta_2)^6*sin(\theta_3)^5*sin(\theta_4)^4 \nonumber \\
&\times &
  sin(\theta_5)^3*sin(\theta_6)^2*sin(\theta_7) ={32 \pi^4 \over 105} \nonumber 
\ea
The angular integrations when averaging using the Cornell potential, are done as in the previous section, with obvious changes
\ba \langle r_{12} \rangle &=& \sqrt{2}\langle |\vec \rho | \rangle  
={\sqrt{2}R_9 \over \Omega_9}
\int d\Omega_9 d_{12}  \nonumber \\
&\approx & 0.773 R_9  \ea
\ba \bigg< {1 \over r_{12}} \bigg> &=&{1 \over R_9 \Omega_9 } \int { d\Omega_9 \over \sqrt{2} d_{12}}  \nonumber \\
&\approx & 1.55/R_9 
  \ea

\section{All-charm tetraquarks}
\subsection{Color-spin wave functions} \label{sec_kinematics_cccc}
The natural way to define the wave function of the $\bar c \bar c cc$
tetraquark, is to start with $cc$ and $\bar c \bar c$ diquarks.
Fermi symmetry requires wave functions to be antisymmetric
under quark permutations, leaving two possibilities
\ba
color &=& [\bar 3] \,\,(antysym), \,\,\, spin=1 (sym) \\
color &=& [6] \,\,(sym), \,\,\, spin=0 \,\, (antisym)
\ea
The total color is zero, and therefore the wave function should take the form
\be \label{eqn_psi_tetra}
\psi=F_{\bar 3 3}(\xi_i) C_{\bar 3 3}(c_i) S_1(s_i)+F_{6 \bar 6}(\xi_i) C_{ 6 \bar 6} (c_i) S_0(s_i)
\ee
where the two spatial wave functions $F$ depend in general on 9 Jacobi coordinates, $C(c_i),i=1..4$, 
on color indices, and $S(s_i),i=1..4$ quark spin variables.
If particles 1,2 are quarks and 3,4 antiquarks,
one can define the color and spin structures as
\ba \label{eqn_diquark_wfs}
C_{\bar 3 3}(c_1 c_2 c_3 c_4) &=& \sum_c \epsilon_{c_1,c_2,c} \epsilon_{c_3,c_4,c}/\sqrt{12} \\
&=&
\big(\delta_{c_1 c_3}\delta_{c_2 c_4}-
\delta_{c_1 c_4}\delta_{c_2 c_3}\big)/\sqrt{12} \nonumber \\
C_{6\bar 6}(c_1 c_2 c_3 c_4) &=& \big(\delta_{c_1 c_3}\delta_{c_2 c_4}+
\delta_{c_1 c_4}\delta_{c_2 c_3}\big)/\sqrt{24} \nonumber \\
S_0 &=& (\uparrow  \downarrow- \downarrow \uparrow)/\sqrt{2} \nonumber \\
S_1 &=& (\uparrow  \downarrow+ \downarrow \uparrow)/\sqrt{2}
\nonumber
\ea
with the normalization
\be \langle \psi | \psi \rangle= \int d^9 \xi |F_{\bar 3 3,6 \bar 6}|^2 \ee
Note that, while the color structures in (\ref{eqn_psi_tetra}) are
not orthogonal to each other, the spin structures are.
So, neglecting the spin-dependent forces (small for charm quarks), 
we expect the Hamiltonian to be diagonal, in terms of $F_{\bar 3 3},F_{6 \bar 6}$.

Following~\cite{Badalian:1985es}, we assume the color dependence of the forces to be similar to those from one-gluon-exchange, namely
\be V=\sum_{i>j} w_{ij}(r_{ij})  \sum_{A=1,8} \frac 14 
\lambda^A_i \lambda^A_j \ee
The lower indices of Gell-Mann matrices indicate on which 
quark color index they act. 

An additional condition is that
for antiquarks (3,4 in current setting),  we have to conjugate $$\lambda^A\rightarrow \tilde \lambda^A= -\lambda^{A*} $$
whereby five contributions  change sign and three do not. 
Averaging $V$ over $\psi$ explicitly, and summing over all indices (we did it using Mathematica) we obtained two  expressions
(which agree with those in
 \cite{Badalian:1985es})
\begin{widetext}
\ba \langle\psi | V | \rangle =  
\begin{bmatrix}
 -{1\over 3} F_{\bar 3 3}^2 (w_{13} + w_{14} + w_{24} + w_{32}+2w_{12}+2w_{34}),  \nonumber \\
 -{1\over 6}  F_{6 \bar 6}^2 \big(-2 w_{12} - 2 w_{34}  
+ 5 w_{24}+5 w_{13}+5 w_{32} +5 w_{14}\big) \nonumber
\end{bmatrix}
\ea 
Note that in the $6\bar 6$ channel, there are
repulsive quark-quark contributions.
Furthermore, $assuming$ that all $w_{ij}$ are the same, one arrives at the conclusion in~\cite{Badalian:1985es}, namely that under these assumptions
both color-spin tetraquark structures have {\em the same} interactions. It  is
 $twice$ stronger than
 in the $Q \bar Q$ channel (while including all six $i,j$  pairs).

\section{Averaging four Wilson lines over the $SU(3)$ rotations}\label{sec_Weingarten}
In this appendix we present expressions for 8 $SU(N_c)$ matrices, averaged over the invariant Haar measure. We do not use them in this paper, since
both for the $qqq$ and $qq \bar q \bar q$ cases considered it turned out
that the color wave functions are such that these complicated formulae
are not needed. Yet they would be needed for four Wilson lines with arbitrary colors. Since we derived such formulae anyway, we decided to show them in
this appendix. 
\subsection{Weingarten-style formula}
One way to carry the color averaging over four pairs of  unitary matrices can be carried using the Weingarten method, producing the following expressions
\bea
\label{WEI}
&&\bigg<U^{a_1}_{c_1}U^{\dagger b_1}_{d_1}U^{a_2}_{c_2}U^{\dagger b_2}_{d_2}U^{a_3}_{c_3}U^{\dagger b_3}_{d_3}
U^{a_4}_{c_4}U^{\dagger b_4}_{d_4}\bigg>_U=\nonumber\\
+&&\frac{6-8N^2_c+N_c^4}{N_c^2(N_c^2-1)(N_c^2-4)(N_c^2-9)}
\sum_{n=1}^{4!}\delta^{a_1a_2a_3a_4}_{P_n(d_1d_2d_3d_4)}\delta^{c_1c_2c_3c_4}_{P_n(b_1b_2b_3 b_4)}\nonumber\\
+&&\frac{4N_c-N_c^3}{N_c^2(N_c^2-1)(N_c^2-4)(N_c^2-9)}
\sum_{n=1}^{4!}\delta^{a_1a_2a_3a_4}_{P_n(d_1d_2d_3d_4)}\nonumber\\
&&\times
\bigg(\delta^{c_1c_2c_3c_4}_{P_n(b_2b_1b_3b_4)}+\delta^{c_1c_2c_3c_4}_{P_n(b_1b_3b_2b_4)}+\delta^{c_1c_2c_3c_4}_{P_n(b_1b_2b_4b_3)}
+\delta^{c_1c_2c_3c_4}_{P_n(b_4b_2b_3b_1)}+\delta^{c_1c_2c_3c_4}_{P_n(b_1b_4b_3b_2)}+\delta^{c_1c_2c_3c_4}_{P_n(b_3b_2b_1b_4)}
\bigg)\nonumber\\
+&&\frac{6+N_c^2}{N_c^2(N_c^2-1)(N_c^2-4)(N_c^2-9)}
\sum_{n=1}^{4!}\delta^{a_1a_2a_3a_4}_{P_n(d_1d_2d_3d_4)}
\bigg(\delta^{c_1c_2c_3c_4}_{P_n(b_2b_1b_4b_3)}+\delta^{c_1c_2c_3c_4}_{P_n(b_3b_4b_1b_2)}+\delta^{c_1c_2c_3c_4}_{P_n(b_4b_3b_2b_1)}
\bigg)\nonumber\\
+&&\frac{-3+2N_c^2}{N_c^2(N_c^2-1)(N_c^2-4)(N_c^2-9)}
\sum_{n=1}^{4!}\delta^{a_1a_2a_3a_4}_{P_n(d_1d_2d_3d_4)}\nonumber\\
&&\times
\bigg(\delta^{c_1c_2c_3c_4}_{P_n(b_2b_3b_1b_4)}+\delta^{c_1c_2c_3c_4}_{P_n(b_1b_3b_4b_2)}+\delta^{c_1c_2c_3c_4}_{P_n(b_3b_2b_4b_1)}
+\delta^{c_1c_2c_3c_4}_{P_n(b_2b_4b_3b_1)}+\delta^{c_1c_2c_3c_4}_{P_n(b_1b_4b_2b_3)}+\delta^{c_1c_2c_3c_4}_{P_n(b_3b_1b_2b_4)}
+\delta^{c_1c_2c_3c_4}_{P_n(b_4b_1b_3b_2)}+\delta^{c_1c_2c_3c_4}_{P_n(b_4b_2b_1b_3)}
\bigg)\nonumber\\
+&&\frac{-5N_c}{N_c^2(N_c^2-1)(N_c^2-4)(N_c^2-9)}
\sum_{n=1}^{4!}\delta^{a_1a_2a_3a_4}_{P_n(d_1d_2d_3d_4)}\nonumber\\
&&\times
\bigg(\delta^{c_1c_2c_3c_4}_{P_n(b_2b_3b_4b_1)}+\delta^{c_1c_2c_3c_4}_{P_n(b_4b_1b_2b_3)}
+\delta^{c_1c_2c_3c_4}_{P_n(b_2b_4b_1b_3)}+\delta^{c_1c_2c_3c_4}_{P_n(b_3b_1b_4b_2)}+\delta^{c_1c_2c_3c_4}_{P_n(b_3b_4b_2b_1)}
+\delta^{c_1c_2c_3c_4}_{P_n(b_4b_3b_1b_2)}
\bigg)\nonumber\\
\eea
In the large $N_c$ limit, it simplifies considerably with the result
\bea
\label{WEI2}
&&\bigg<U^{a_1}_{c_1}U^{\dagger b_1}_{d_1}U^{a_2}_{c_2}U^{\dagger b_2}_{d_2}U^{a_3}_{c_3}U^{\dagger b_3}_{d_3}
U^{a_4}_{c_4}U^{\dagger b_4}_{d_4}\bigg>_U=
\frac 1{N_c^4}
\sum_{n=1}^{4!}\delta^{a_1a_2a_3a_4}_{P_n(d_1d_2d_3d_4)}\delta^{c_1c_2c_3c_4}_{P_n(b_1b_2b_3 b_4)}
+{\cal O}\bigg(\frac 1{N_c^5}\bigg)
\eea
We are using  the shorthand notation 
\bea
\sum_{n=1}^{4!}\delta^{a_1a_2a_3a_4}_{P_n(d_1d_2d_3d_4)}\delta^{c_1c_2c_3c_4}_{P_n(b_1b_2b_3 b_4)}=
\bigg(\delta^{a_1}_{d_1}\delta^{a_2}_{d_2}\delta^{a_3}_{d_3}\delta^{a_4}_{d_4}
\delta^{c_1}_{b_1}\delta^{c_2}_{b_2}\delta^{c_3}_{b_3}\delta^{c_4}_{b_4}
+{\rm perm.}\bigg)
\eea 
\end{widetext}
where the sum is over the 4!  $P_n$ elements of the permutation group $S_4$, which are explicitly
\bea
P_1(a_1a_2a_3a_4)&=&(a_1a_2a_3a_4)\nonumber\\
P_2(a_1a_2a_3a_4)&=&(a_1a_2a_4a_3)\nonumber\\
P_3(a_1a_2a_3a_4)&=&(a_1a_3a_2a_4)\nonumber\\
P_4(a_1a_2a_3a_4)&=&(a_1a_3a_4a_2)\nonumber\\
P_5(a_1a_2a_3a_4)&=&(a_1a_4a_2a_3)\nonumber\\
P_6(a_1a_2a_3a_4)&=&(a_1a_2a_3a_4)\nonumber\\
P_7(a_1a_2a_3a_4)&=&(a_1a_4a_3a_2) \nonumber
\eea
\bea
P_8(a_1a_2a_3a_4)&=&(a_2a_1a_3a_4)\nonumber\\
P_9(a_1a_2a_3a_4)&=&(a_2a_1a_4a_3)\nonumber\\
P_{10}(a_1a_2a_3a_4)&=&(a_2a_3a_1a_4)\nonumber\\
P_{11}(a_1a_2a_3a_4)&=&(a_2a_3a_4a_1)\nonumber\\
P_{12}(a_1a_2a_3a_4)&=&(a_2a_4a_1a_3)\nonumber\\
P_{13}(a_1a_2a_3a_4)&=&(a_2a_4a_3a_1)  \nonumber
\eea
\bea
P_{14}(a_1a_2a_3a_4)&=&(a_3a_1a_2a_4)\nonumber\\
P_{15}(a_1a_2a_3a_4)&=&(a_3a_1a_4a_2)\nonumber\\
P_{16}(a_1a_2a_3a_4)&=&(a_3a_2a_4a_1)\nonumber\\
P_{17}(a_1a_2a_3a_4)&=&(a_3a_4a_1a_2)\nonumber\\
P_{18}(a_1a_2a_3a_4)&=&(a_3a_4a_2a_1)  \nonumber
\eea
\bea
P_{19}(a_1a_2a_3a_4)&=&(a_4a_1a_2a_3)\nonumber\\
P_{20}(a_1a_2a_3a_4)&=&(a_4a_1a_3a_2)\nonumber\\
P_{21}(a_1a_2a_3a_4)&=&(a_4a_2a_1a_3)\nonumber\\
P_{22}(a_1a_2a_3a_4)&=&(a_4a_2a_3a_1)\nonumber\\
P_{23}(a_1a_2a_3a_4)&=&(a_4a_3a_1a_2)\nonumber\\
P_{24}(a_1a_2a_3a_4)&=&(a_4a_3a_2a_1)
\eea
Note that coefficients have different powers of $1/N_c$, so when
$N_c\rightarrow \infty$ the only one left is the first raw. 
However at $N_c=3$ they are all identical (by modulus) and have alternating signs, so they conspire to cancel. At the other hand, coefficients have
$1/(N_c^2-9)$ factors prohibiting direct usage of this expressions when  $N_c=3$ (or smaller), without investigations of those cancellations.
  Clearly, the result (\ref{WEI})
is useful for numerical implementation for $N_c>3$. 
For $N_c\leq p=4$ we now provide alternative  and finite identities.

\subsection{Creutz formula}
Another way to carry the color averaging in (\ref{WEI})
which is pole free, is by determinantal reduction. More specifically, we will use 
Creutz's identities~\cite{Creutz:1978ub}
\be
\label{DET1}
\bigg< \prod_{i=1}^{N_c}U^{a_i}_{c_i}\bigg>_U =
\frac 1{N_c!}\epsilon^{a_1...a_{N_c}}\epsilon_{c_1 ...c_{N_c}}
\ee
and
\bea
\label{DET2}
U^{\dagger a}_c=&&\frac 1{(N_c-1)!}\nonumber\\
&&\times\epsilon^{aa_1...a_{N_c-1}}
\epsilon_{cc_1 ...c_{N_c-1}}
U^{a_1}_{c_1}...U^{a_{N_c-1}}_{c_{{N_c-1}}}\nonumber\\
\eea
where $\epsilon^{a_1...a_{N_c}}$ is the Levi-Cevita tensor of rank-$N_c$, with $\epsilon^{1...N_c}=1$.

With this in mind, we can substitute (\ref{DET2}) for the $p=4$ string of $U^\dagger$ in (\ref{WEI}), with the result

\begin{widetext}
\bea
\label{DET3}
&&\bigg<U^{a_1}_{c_1}U^{\dagger b_1}_{d_1}U^{a_2}_{c_2}U^{\dagger b_2}_{d_2}U^{a_3}_{c_3}U^{\dagger b_3}_{d_3}
U^{a_4}_{c_4}U^{\dagger b_4}_{d_4}\bigg>_U=\nonumber\\
&&\frac 1{((N_c-1)!)^{p=4}}
\prod^{p=4}_{m=1} 
\epsilon^{b_mb_{1m}...b_{(N_c-1)m}}
\epsilon_{d_md_{1m} ..._{(N_c-1)m}}
\bigg<\prod_{m=1}^{p=4}
U^{a_m}_{c_m} U^{b_{1m}}_{d_{1m}}...
U^{b_{(N_c-1)m}}_{d_{(N_c-1)m}}
\bigg>_U
\eea
Each of the product is composed rank-$N_c$ Levi-Cevita tensors. 
The last unitary averaging in (\ref{DET3}) can be undone by  the determinantal identity (\ref{DET1}) 
\bea
\label{DET4}
&&\bigg<U^{a_1}_{c_1}U^{\dagger b_1}_{d_1}U^{a_2}_{c_2}U^{\dagger b_2}_{d_2}U^{a_3}_{c_3}U^{\dagger b_3}_{d_3}
U^{a_4}_{c_4}U^{\dagger b_4}_{d_4}\bigg>_U=
\bigg(\frac 1{((N_c-1)!)^p}\frac{2!... (N_c-1)!}{(p+1)!...(p+N_c-1)!}\bigg)_{p=4}
\nonumber\\
&&\prod^{p=4}_{m=1} 
\epsilon^{a_mb_{1m}...b_{(N_c-1)m}}
\epsilon_{c_md_{1m} ...d_{(N_c-1)m}}
\bigg(\prod_{m=1}^{p=4}
\epsilon^{a_m b_{1m}...b_{(N_c-1)m}}
\epsilon_{c_m d_{1m}...d_{(N_c-1)m}}
+{\rm perm.}\bigg)
\eea
\end{widetext}
The combinatorical pre-factor follows from the fact that the group integration is invariant under the permutation of the U-factors, with
a total of $(pN_c)!$ permutations. Note that many permutations are identical, with only
$\frac {(pN_c)!}{(p!(N_c!)^p)}$ independent ones.
Clearly all coefficients in (\ref{DET4}) are finite for 
any value of $N_c$, in contrast to (\ref{WEI}). We note that (\ref{DET4}) holds for the averaging of even higher
products $(UU^\dagger)^p$, and is free
of poles whatever $p,N_c$.

\section{CNZ formula}
Finally, another alternative formula for the color averaging can be obtained by
using the graphical color projection rules developed in CNZ paper~\cite{Chernyshev:1995gj}. More specifically, 
the color averaging $(UU^\dagger)^4$ with $p=4$, can be tied recursively to $p=3$ by reduction
\begin{widetext}
\bea
\label{DET6}
&&\bigg<U^{a_1}_{c_1}U^{\dagger b_1}_{d_1}U^{a_2}_{c_2}U^{\dagger b_2}_{d_2}U^{a_3}_{c_3}U^{\dagger b_3}_{d_3}
U^{a_4}_{c_4}U^{\dagger b_4}_{d_4}\bigg>_U=\nonumber\\
&&+\bigg(\bigg[\frac 1{N_c}\delta^{a_1}_{d_1}\delta^{b_1}_{c_1}\,{\bf 1}_1\bigg]
\bigg<U^{a_2}_{c_2}U^{\dagger b_2}_{d_2}U^{a_3}_{c_3}U^{\dagger b_3}_{d_3}
U^{a_4}_{c_4}U^{\dagger b_4}_{d_4}\bigg>_U+{\rm perm.}\bigg)\nonumber\\
&& +\bigg(\big([\lambda^A_1]^{a_1}_{d_1}[\lambda^B_2]^{a_2}_{d_2}[\lambda^C_3]^{a_3}_{d_3}[\lambda^D_4]^{a_4}_{d_4}\big)\,
\big([\lambda^I_1]^{b_1}_{c_1}[\lambda^J_2]^{b_2}_{c_2}[\lambda^K_3]^{b_3}_{c_3}[\lambda^L_4]^{b_4}_{c_4}\big)\,X^{ABCD}\,X^{IJKL}
\bigg)
\eea
with the color factors
\bea
X^{ABCD}=&&\frac 1{16{(N_c^2-1)^2}}(\delta^{AB}\delta^{CD}+
\delta^{AC}\delta^{BD}+\delta^{AD}\delta^{BC})\nonumber\\
&&+\frac 1{4(N_c^2-1)}\frac {N_c^2}{4{(N_c^2-4)^2}}
\big(d^{ABM}d^{CDM}+d^{ACM}d^{BDM}+d^{ADM}d^{BCM}\big)\nonumber\\
&&+\frac 1{4(N_c^2-1)}\frac 1{4N_c^2}
\big(f^{ABM}f^{CDM}+f^{ACM}f^{BDM}+f^{ADM}f^{BCM}\big)\nonumber\\
&&+\frac 1{4(N_c^2-1)}\frac {1}{4{(N_c^2-4)}}
\big(d^{ABM}f^{CDM}+d^{ACM}f^{BDM}+d^{ADM}f^{BCM}\big)\nonumber\\
&&+\frac 1{4(N_c^2-1)}\frac {1}{4{(N_c^2-4)}}
\big(f^{ABM}d^{CDM}+f^{ACM}d^{BDM}+f^{ADM}d^{BCM}\big)
\eea
and similarly for $X^{IJKL}$.
The reduced averaging over $(UU^\dagger)^3$ gives
\bea
\label{V3}
&&\bigg<U^{a_2}_{c_2}U^{\dagger b_2}_{d_2}U^{a_3}_{c_3}U^{\dagger b_3}_{d_3}
U^{a_4}_{c_4}U^{\dagger b_4}_{d_4}\bigg>_U=\bigg[\frac 1{N_c}\delta^{a_2}_{d_2}\delta^{b_2}_{c_2}\,{\bf 1}_2\bigg]
\bigg[\frac 1{N_c}\delta^{a_3}_{d_3}\delta^{b_3}_{c_3}\,{\bf 1}_3\bigg]
\bigg[\frac 1{N_c}\delta^{a_4}_{d_4}\delta^{b_4}_{c_4}\,{\bf 1}_4\bigg]
\nonumber\\
&&+\bigg(\bigg[\frac 1{N_c}\delta^{a_2}_{d_2}\delta^{b_2}_{c_2}\,{\bf 1}_2\bigg]
\bigg[\frac 1{4(N_c^2-1)}[\lambda_3^A]^{a_3}_{d_3}[\lambda_4^A]^{a_4}_{d_4}[\lambda_3^B]^{b_3}_{c_3}[\lambda_4^B]^{b_4}_{c_4}\bigg]+2\,\,{\rm  perm.}\bigg)\nonumber\\
&&+\frac{N_c}{8(N_c^2-1)(N_c^2-4)}
\bigg[d^{ABC}[\lambda_2^A]^{a_2}_{d_2}
[\lambda_3^B]^{a_3}_{d_3}
[\lambda_4^C]^{a_4}_{d_4}\bigg]
\bigg[d^{IJK}[\lambda_2^I]^{b_2}_{c_2}
[\lambda_3^J]^{b_3}_{c_3}
[\lambda_4^K]^{b_4}_{c_4}\bigg]
\nonumber\\
&&+\frac{1}{8N_c(N_c^2-1)}
\bigg[f^{ABC}[\lambda_2^A]^{a_2}_{d_2}
[\lambda_3^B]^{a_3}_{d_3}
[\lambda_4^C]^{a_4}_{d_4}\bigg]
\bigg[f^{IJK}[\lambda_2^I]^{b_2}_{c_2}
[\lambda_3^J]^{b_3}_{c_3}
[\lambda_4^K]^{b_4}_{c_4}\bigg]
\eea

\end{widetext}

\bibliography{temp}

\begin{thebibliography}{26}%
\makeatletter
\providecommand \@ifxundefined [1]{%
 \@ifx{#1\undefined}
}%
\providecommand \@ifnum [1]{%
 \ifnum #1\expandafter \@firstoftwo
 \else \expandafter \@secondoftwo
 \fi
}%
\providecommand \@ifx [1]{%
 \ifx #1\expandafter \@firstoftwo
 \else \expandafter \@secondoftwo
 \fi
}%
\providecommand \natexlab [1]{#1}%
\providecommand \enquote  [1]{``#1''}%
\providecommand \bibnamefont  [1]{#1}%
\providecommand \bibfnamefont [1]{#1}%
\providecommand \citenamefont [1]{#1}%
\providecommand \href@noop [0]{\@secondoftwo}%
\providecommand \href [0]{\begingroup \@sanitize@url \@href}%
\providecommand \@href[1]{\@@startlink{#1}\@@href}%
\providecommand \@@href[1]{\endgroup#1\@@endlink}%
\providecommand \@sanitize@url [0]{\catcode `\\12\catcode `\$12\catcode
  `\&12\catcode `\#12\catcode `\^12\catcode `\_12\catcode `\%12\relax}%
\providecommand \@@startlink[1]{}%
\providecommand \@@endlink[0]{}%
\providecommand \url  [0]{\begingroup\@sanitize@url \@url }%
\providecommand \@url [1]{\endgroup\@href {#1}{\urlprefix }}%
\providecommand \urlprefix  [0]{URL }%
\providecommand \Eprint [0]{\href }%
\providecommand \doibase [0]{http://dx.doi.org/}%
\providecommand \selectlanguage [0]{\@gobble}%
\providecommand \bibinfo  [0]{\@secondoftwo}%
\providecommand \bibfield  [0]{\@secondoftwo}%
\providecommand \translation [1]{[#1]}%
\providecommand \BibitemOpen [0]{}%
\providecommand \bibitemStop [0]{}%
\providecommand \bibitemNoStop [0]{.\EOS\space}%
\providecommand \EOS [0]{\spacefactor3000\relax}%
\providecommand \BibitemShut  [1]{\csname bibitem#1\endcsname}%
\let\auto@bib@innerbib\@empty
\bibitem [{\citenamefont {Shuryak}\ and\ \citenamefont
  {Zahed}(2023{\natexlab{a}})}]{Shuryak:2021fsu}%
  \BibitemOpen
  \bibfield  {author} {\bibinfo {author} {\bibfnamefont {E.}~\bibnamefont
  {Shuryak}}\ and\ \bibinfo {author} {\bibfnamefont {I.}~\bibnamefont
  {Zahed}},\ }\href {\doibase 10.1103/PhysRevD.107.034023} {\bibfield
  {journal} {\bibinfo  {journal} {Phys. Rev. D}\ }\textbf {\bibinfo {volume}
  {107}},\ \bibinfo {pages} {034023} (\bibinfo {year} {2023}{\natexlab{a}})},\
  \Eprint {http://arxiv.org/abs/2110.15927} {arXiv:2110.15927 [hep-ph]}
  \BibitemShut {NoStop}%
\bibitem [{\citenamefont {Simonov}(1966)}]{Simonov:1965ei}%
  \BibitemOpen
  \bibfield  {author} {\bibinfo {author} {\bibfnamefont {Y.~A.}\ \bibnamefont
  {Simonov}},\ }\href@noop {} {\bibfield  {journal} {\bibinfo  {journal} {Sov.
  J. Nucl. Phys.}\ }\textbf {\bibinfo {volume} {3}},\ \bibinfo {pages} {461}
  (\bibinfo {year} {1966})}\BibitemShut {NoStop}%
\bibitem [{\citenamefont {Shuryak}\ and\ \citenamefont
  {Torres-Rincon}(2019)}]{Shuryak:2018lgd}%
  \BibitemOpen
  \bibfield  {author} {\bibinfo {author} {\bibfnamefont {E.}~\bibnamefont
  {Shuryak}}\ and\ \bibinfo {author} {\bibfnamefont {J.~M.}\ \bibnamefont
  {Torres-Rincon}},\ }\href {\doibase 10.1103/PhysRevC.100.024903} {\bibfield
  {journal} {\bibinfo  {journal} {Phys. Rev. C}\ }\textbf {\bibinfo {volume}
  {100}},\ \bibinfo {pages} {024903} (\bibinfo {year} {2019})},\ \Eprint
  {http://arxiv.org/abs/1805.04444} {arXiv:1805.04444 [hep-ph]} \BibitemShut
  {NoStop}%
\bibitem [{\citenamefont {Badalian}\ \emph {et~al.}(1987)\citenamefont
  {Badalian}, \citenamefont {Ioffe},\ and\ \citenamefont
  {Smilga}}]{Badalian:1985es}%
  \BibitemOpen
  \bibfield  {author} {\bibinfo {author} {\bibfnamefont {A.~M.}\ \bibnamefont
  {Badalian}}, \bibinfo {author} {\bibfnamefont {B.~L.}\ \bibnamefont {Ioffe}},
  \ and\ \bibinfo {author} {\bibfnamefont {A.~V.}\ \bibnamefont {Smilga}},\
  }\href {\doibase 10.1016/0550-3213(87)90248-3} {\bibfield  {journal}
  {\bibinfo  {journal} {Nucl. Phys. B}\ }\textbf {\bibinfo {volume} {281}},\
  \bibinfo {pages} {85} (\bibinfo {year} {1987})}\BibitemShut {NoStop}%
\bibitem [{\citenamefont {Chernyshev}\ \emph {et~al.}(1996)\citenamefont
  {Chernyshev}, \citenamefont {Nowak},\ and\ \citenamefont
  {Zahed}}]{Chernyshev:1995gj}%
  \BibitemOpen
  \bibfield  {author} {\bibinfo {author} {\bibfnamefont {S.}~\bibnamefont
  {Chernyshev}}, \bibinfo {author} {\bibfnamefont {M.~A.}\ \bibnamefont
  {Nowak}}, \ and\ \bibinfo {author} {\bibfnamefont {I.}~\bibnamefont
  {Zahed}},\ }\href {\doibase 10.1103/PhysRevD.53.5176} {\bibfield  {journal}
  {\bibinfo  {journal} {Phys. Rev. D}\ }\textbf {\bibinfo {volume} {53}},\
  \bibinfo {pages} {5176} (\bibinfo {year} {1996})},\ \Eprint
  {http://arxiv.org/abs/hep-ph/9510326} {arXiv:hep-ph/9510326} \BibitemShut
  {NoStop}%
\bibitem [{\citenamefont {Ferraris}\ \emph {et~al.}(1995)\citenamefont
  {Ferraris}, \citenamefont {Giannini}, \citenamefont {Pizzo}, \citenamefont
  {Santopinto},\ and\ \citenamefont {Tiator}}]{Ferraris:1995ui}%
  \BibitemOpen
  \bibfield  {author} {\bibinfo {author} {\bibfnamefont {M.}~\bibnamefont
  {Ferraris}}, \bibinfo {author} {\bibfnamefont {M.~M.}\ \bibnamefont
  {Giannini}}, \bibinfo {author} {\bibfnamefont {M.}~\bibnamefont {Pizzo}},
  \bibinfo {author} {\bibfnamefont {E.}~\bibnamefont {Santopinto}}, \ and\
  \bibinfo {author} {\bibfnamefont {L.}~\bibnamefont {Tiator}},\ }\href
  {\doibase 10.1016/0370-2693(95)01091-2} {\bibfield  {journal} {\bibinfo
  {journal} {Phys. Lett. B}\ }\textbf {\bibinfo {volume} {364}},\ \bibinfo
  {pages} {231} (\bibinfo {year} {1995})}\BibitemShut {NoStop}%
\bibitem [{\citenamefont {Gandhi}\ \emph {et~al.}(2018)\citenamefont {Gandhi},
  \citenamefont {Shah},\ and\ \citenamefont {Rai}}]{Gandhi:2018lez}%
  \BibitemOpen
  \bibfield  {author} {\bibinfo {author} {\bibfnamefont {K.}~\bibnamefont
  {Gandhi}}, \bibinfo {author} {\bibfnamefont {Z.}~\bibnamefont {Shah}}, \ and\
  \bibinfo {author} {\bibfnamefont {A.~K.}\ \bibnamefont {Rai}},\ }\href
  {\doibase 10.1140/epjp/i2018-12318-1} {\bibfield  {journal} {\bibinfo
  {journal} {Eur. Phys. J. Plus}\ }\textbf {\bibinfo {volume} {133}},\ \bibinfo
  {pages} {512} (\bibinfo {year} {2018})},\ \Eprint
  {http://arxiv.org/abs/1811.00251} {arXiv:1811.00251 [hep-ph]} \BibitemShut
  {NoStop}%
\bibitem [{\citenamefont {Brambilla}\ \emph {et~al.}(2022)\citenamefont
  {Brambilla} \emph {et~al.}}]{Brambilla:2022ura}%
  \BibitemOpen
  \bibfield  {author} {\bibinfo {author} {\bibfnamefont {N.}~\bibnamefont
  {Brambilla}} \emph {et~al.},\ }\href@noop {} {\  (\bibinfo {year} {2022})},\
  \Eprint {http://arxiv.org/abs/2203.16583} {arXiv:2203.16583 [hep-ph]}
  \BibitemShut {NoStop}%
\bibitem [{\citenamefont {Koma}\ and\ \citenamefont
  {Koma}(2017)}]{Koma:2017hcm}%
  \BibitemOpen
  \bibfield  {author} {\bibinfo {author} {\bibfnamefont {Y.}~\bibnamefont
  {Koma}}\ and\ \bibinfo {author} {\bibfnamefont {M.}~\bibnamefont {Koma}},\
  }\href {\doibase 10.1103/PhysRevD.95.094513} {\bibfield  {journal} {\bibinfo
  {journal} {Phys. Rev. D}\ }\textbf {\bibinfo {volume} {95}},\ \bibinfo
  {pages} {094513} (\bibinfo {year} {2017})},\ \Eprint
  {http://arxiv.org/abs/1703.06247} {arXiv:1703.06247 [hep-lat]} \BibitemShut
  {NoStop}%
\bibitem [{\citenamefont {Bicudo}\ \emph {et~al.}(2017)\citenamefont {Bicudo},
  \citenamefont {Cardoso}, \citenamefont {Oliveira},\ and\ \citenamefont
  {Silva}}]{1702.07789}%
  \BibitemOpen
  \bibfield  {author} {\bibinfo {author} {\bibfnamefont {P.}~\bibnamefont
  {Bicudo}}, \bibinfo {author} {\bibfnamefont {M.}~\bibnamefont {Cardoso}},
  \bibinfo {author} {\bibfnamefont {O.}~\bibnamefont {Oliveira}}, \ and\
  \bibinfo {author} {\bibfnamefont {P.~J.}\ \bibnamefont {Silva}},\ }\href
  {\doibase 10.1103/PhysRevD.96.074508} {\bibfield  {journal} {\bibinfo
  {journal} {Phys. Rev. D}\ }\textbf {\bibinfo {volume} {96}},\ \bibinfo
  {pages} {074508} (\bibinfo {year} {2017})},\ \Eprint
  {http://arxiv.org/abs/1702.07789} {arXiv:1702.07789 [hep-lat]} \BibitemShut
  {NoStop}%
\bibitem [{\citenamefont {Shuryak}(1982)}]{Shuryak:1981ff}%
  \BibitemOpen
  \bibfield  {author} {\bibinfo {author} {\bibfnamefont {E.~V.}\ \bibnamefont
  {Shuryak}},\ }\href {\doibase 10.1016/0550-3213(82)90478-3} {\bibfield
  {journal} {\bibinfo  {journal} {Nucl. Phys. B}\ }\textbf {\bibinfo {volume}
  {203}},\ \bibinfo {pages} {93} (\bibinfo {year} {1982})}\BibitemShut
  {NoStop}%
\bibitem [{\citenamefont {Sch\"afer}\ and\ \citenamefont
  {Shuryak}(1998)}]{Schafer:1996wv}%
  \BibitemOpen
  \bibfield  {author} {\bibinfo {author} {\bibfnamefont {T.}~\bibnamefont
  {Sch\"afer}}\ and\ \bibinfo {author} {\bibfnamefont {E.~V.}\ \bibnamefont
  {Shuryak}},\ }\href {\doibase 10.1103/RevModPhys.70.323} {\bibfield
  {journal} {\bibinfo  {journal} {Rev. Mod. Phys.}\ }\textbf {\bibinfo {volume}
  {70}},\ \bibinfo {pages} {323} (\bibinfo {year} {1998})},\ \Eprint
  {http://arxiv.org/abs/hep-ph/9610451} {arXiv:hep-ph/9610451} \BibitemShut
  {NoStop}%
\bibitem [{\citenamefont {Workman}\ \emph {et~al.}(2022)\citenamefont {Workman}
  \emph {et~al.}}]{ParticleDataGroup:2022pth}%
  \BibitemOpen
  \bibfield  {author} {\bibinfo {author} {\bibfnamefont {R.~L.}\ \bibnamefont
  {Workman}} \emph {et~al.} (\bibinfo {collaboration} {Particle Data Group}),\
  }\href {\doibase 10.1093/ptep/ptac097} {\bibfield  {journal} {\bibinfo
  {journal} {PTEP}\ }\textbf {\bibinfo {volume} {2022}},\ \bibinfo {pages}
  {083C01} (\bibinfo {year} {2022})}\BibitemShut {NoStop}%
\bibitem [{\citenamefont {Grosse}\ and\ \citenamefont
  {Martin}(1980)}]{Grosse:1979xm}%
  \BibitemOpen
  \bibfield  {author} {\bibinfo {author} {\bibfnamefont {H.}~\bibnamefont
  {Grosse}}\ and\ \bibinfo {author} {\bibfnamefont {A.}~\bibnamefont
  {Martin}},\ }\href {\doibase 10.1016/0370-1573(80)90031-9} {\bibfield
  {journal} {\bibinfo  {journal} {Phys. Rept.}\ }\textbf {\bibinfo {volume}
  {60}},\ \bibinfo {pages} {341} (\bibinfo {year} {1980})}\BibitemShut
  {NoStop}%
\bibitem [{\citenamefont {Callan}\ \emph {et~al.}(1978)\citenamefont {Callan},
  \citenamefont {Dashen}, \citenamefont {Gross}, \citenamefont {Wilczek},\ and\
  \citenamefont {Zee}}]{Callan:1978ye}%
  \BibitemOpen
  \bibfield  {author} {\bibinfo {author} {\bibfnamefont {C.~G.}\ \bibnamefont
  {Callan}, \bibfnamefont {Jr.}}, \bibinfo {author} {\bibfnamefont {R.~F.}\
  \bibnamefont {Dashen}}, \bibinfo {author} {\bibfnamefont {D.~J.}\
  \bibnamefont {Gross}}, \bibinfo {author} {\bibfnamefont {F.}~\bibnamefont
  {Wilczek}}, \ and\ \bibinfo {author} {\bibfnamefont {A.}~\bibnamefont
  {Zee}},\ }\href {\doibase 10.1103/PhysRevD.18.4684} {\bibfield  {journal}
  {\bibinfo  {journal} {Phys. Rev. D}\ }\textbf {\bibinfo {volume} {18}},\
  \bibinfo {pages} {4684} (\bibinfo {year} {1978})}\BibitemShut {NoStop}%
\bibitem [{\citenamefont {Shuryak}\ and\ \citenamefont
  {Zahed}(2023{\natexlab{b}})}]{Shuryak:2021hng}%
  \BibitemOpen
  \bibfield  {author} {\bibinfo {author} {\bibfnamefont {E.}~\bibnamefont
  {Shuryak}}\ and\ \bibinfo {author} {\bibfnamefont {I.}~\bibnamefont
  {Zahed}},\ }\href {\doibase 10.1103/PhysRevD.107.034024} {\bibfield
  {journal} {\bibinfo  {journal} {Phys. Rev. D}\ }\textbf {\bibinfo {volume}
  {107}},\ \bibinfo {pages} {034024} (\bibinfo {year} {2023}{\natexlab{b}})},\
  \Eprint {http://arxiv.org/abs/2111.01775} {arXiv:2111.01775 [hep-ph]}
  \BibitemShut {NoStop}%
\bibitem [{\citenamefont {Shuryak}\ and\ \citenamefont
  {Zahed}(2021)}]{Shuryak:2021mlh}%
  \BibitemOpen
  \bibfield  {author} {\bibinfo {author} {\bibfnamefont {E.}~\bibnamefont
  {Shuryak}}\ and\ \bibinfo {author} {\bibfnamefont {I.}~\bibnamefont
  {Zahed}},\ }\href@noop {} {\  (\bibinfo {year} {2021})},\ \Eprint
  {http://arxiv.org/abs/2112.15586} {arXiv:2112.15586 [hep-ph]} \BibitemShut
  {NoStop}%
\bibitem [{\citenamefont {Guimar\~aes}\ \emph {et~al.}(1981)\citenamefont
  {Guimar\~aes}, \citenamefont {Coelho},\ and\ \citenamefont
  {Chanda}}]{PhysRevD.24.1343}%
  \BibitemOpen
  \bibfield  {author} {\bibinfo {author} {\bibfnamefont {A.~B.}\ \bibnamefont
  {Guimar\~aes}}, \bibinfo {author} {\bibfnamefont {H.~T.}\ \bibnamefont
  {Coelho}}, \ and\ \bibinfo {author} {\bibfnamefont {R.}~\bibnamefont
  {Chanda}},\ }\href {\doibase 10.1103/PhysRevD.24.1343} {\bibfield  {journal}
  {\bibinfo  {journal} {Phys. Rev. D}\ }\textbf {\bibinfo {volume} {24}},\
  \bibinfo {pages} {1343} (\bibinfo {year} {1981})}\BibitemShut {NoStop}%
\bibitem [{\citenamefont {Aaij}\ \emph {et~al.}(2020)\citenamefont {Aaij} \emph
  {et~al.}}]{2006.16957}%
  \BibitemOpen
  \bibfield  {author} {\bibinfo {author} {\bibfnamefont {R.}~\bibnamefont
  {Aaij}} \emph {et~al.} (\bibinfo {collaboration} {LHCb}),\ }\href {\doibase
  10.1016/j.scib.2020.08.032} {\bibfield  {journal} {\bibinfo  {journal} {Sci.
  Bull.}\ }\textbf {\bibinfo {volume} {65}},\ \bibinfo {pages} {1983} (\bibinfo
  {year} {2020})},\ \Eprint {http://arxiv.org/abs/2006.16957} {arXiv:2006.16957
  [hep-ex]} \BibitemShut {NoStop}%
\bibitem [{\citenamefont {Zhang}\ and\ \citenamefont {Yi}(2022)}]{2212.00504}%
  \BibitemOpen
  \bibfield  {author} {\bibinfo {author} {\bibfnamefont {J.}~\bibnamefont
  {Zhang}}\ and\ \bibinfo {author} {\bibfnamefont {K.}~\bibnamefont {Yi}}
  (\bibinfo {collaboration} {CMS}),\ }\href {\doibase 10.22323/1.414.0775}
  {\bibfield  {journal} {\bibinfo  {journal} {PoS}\ }\textbf {\bibinfo {volume}
  {ICHEP2022}},\ \bibinfo {pages} {775} (\bibinfo {year} {2022})},\ \Eprint
  {http://arxiv.org/abs/2212.00504} {arXiv:2212.00504 [hep-ex]} \BibitemShut
  {NoStop}%
\bibitem [{\citenamefont {Xu}(2023)}]{2209.12173}%
  \BibitemOpen
  \bibfield  {author} {\bibinfo {author} {\bibfnamefont {Y.}~\bibnamefont {Xu}}
  (\bibinfo {collaboration} {ATLAS}),\ }\href {\doibase
  10.5506/APhysPolBSupp.16.3-A21} {\bibfield  {journal} {\bibinfo  {journal}
  {Acta Phys. Polon. Supp.}\ }\textbf {\bibinfo {volume} {16}},\ \bibinfo
  {pages} {21} (\bibinfo {year} {2023})},\ \Eprint
  {http://arxiv.org/abs/2209.12173} {arXiv:2209.12173 [hep-ex]} \BibitemShut
  {NoStop}%
\bibitem [{\citenamefont {Karliner}\ and\ \citenamefont
  {Rosner}(2020)}]{2009.04429}%
  \BibitemOpen
  \bibfield  {author} {\bibinfo {author} {\bibfnamefont {M.}~\bibnamefont
  {Karliner}}\ and\ \bibinfo {author} {\bibfnamefont {J.~L.}\ \bibnamefont
  {Rosner}},\ }\href {\doibase 10.1103/PhysRevD.102.114039} {\bibfield
  {journal} {\bibinfo  {journal} {Phys. Rev. D}\ }\textbf {\bibinfo {volume}
  {102}},\ \bibinfo {pages} {114039} (\bibinfo {year} {2020})},\ \Eprint
  {http://arxiv.org/abs/2009.04429} {arXiv:2009.04429 [hep-ph]} \BibitemShut
  {NoStop}%
\bibitem [{\citenamefont {Giron}\ and\ \citenamefont
  {Lebed}(2020)}]{2008.01631}%
  \BibitemOpen
  \bibfield  {author} {\bibinfo {author} {\bibfnamefont {J.~F.}\ \bibnamefont
  {Giron}}\ and\ \bibinfo {author} {\bibfnamefont {R.~F.}\ \bibnamefont
  {Lebed}},\ }\href {\doibase 10.1103/PhysRevD.102.074003} {\bibfield
  {journal} {\bibinfo  {journal} {Phys. Rev. D}\ }\textbf {\bibinfo {volume}
  {102}},\ \bibinfo {pages} {074003} (\bibinfo {year} {2020})},\ \Eprint
  {http://arxiv.org/abs/2008.01631} {arXiv:2008.01631 [hep-ph]} \BibitemShut
  {NoStop}%
\bibitem [{\citenamefont {Badalian}(2023)}]{Badalian:2023krq}%
  \BibitemOpen
  \bibfield  {author} {\bibinfo {author} {\bibfnamefont {A.~M.}\ \bibnamefont
  {Badalian}},\ }\href@noop {} {\  (\bibinfo {year} {2023})},\ \Eprint
  {http://arxiv.org/abs/2305.04585} {arXiv:2305.04585 [hep-ph]} \BibitemShut
  {NoStop}%
\bibitem [{\citenamefont {{Fabre de la Ripelle}}\ and\ \citenamefont
  {Navarro}(1979)}]{FABREDELARIPELLE1979185}%
  \BibitemOpen
  \bibfield  {author} {\bibinfo {author} {\bibfnamefont {M.}~\bibnamefont
  {{Fabre de la Ripelle}}}\ and\ \bibinfo {author} {\bibfnamefont
  {J.}~\bibnamefont {Navarro}},\ }\href {\doibase
  https://doi.org/10.1016/0003-4916(79)90270-7} {\bibfield  {journal} {\bibinfo
   {journal} {Annals of Physics}\ }\textbf {\bibinfo {volume} {123}},\ \bibinfo
  {pages} {185} (\bibinfo {year} {1979})}\BibitemShut {NoStop}%
\bibitem [{\citenamefont {Creutz}(1978)}]{Creutz:1978ub}%
  \BibitemOpen
  \bibfield  {author} {\bibinfo {author} {\bibfnamefont {M.}~\bibnamefont
  {Creutz}},\ }\href {\doibase 10.1063/1.523581} {\bibfield  {journal}
  {\bibinfo  {journal} {J. Math. Phys.}\ }\textbf {\bibinfo {volume} {19}},\
  \bibinfo {pages} {2043} (\bibinfo {year} {1978})}\BibitemShut {NoStop}%
\end{thebibliography}%
\end{document}